\documentclass[prd,aps,twocolumn,floats,floatfix,nofootinbib]{revtex4-1}
\usepackage{amssymb,bm}
\usepackage{graphicx}
\usepackage{dcolumn}
\usepackage{bm}
\usepackage{graphics}
\usepackage{slashed}
\usepackage{enumerate}
\usepackage{amsmath}
\usepackage{hyperref}
\usepackage{url}
\usepackage[usenames,dvipsnames,svgnames,table]{xcolor}
\usepackage{color}
\usepackage{xcolor}
\usepackage{grffile}
\usepackage{booktabs}
\usepackage[toc,page]{appendix}
\usepackage{float}
\usepackage{soul}
\usepackage{subfigure}
\usepackage{multirow}
\usepackage{booktabs}
\usepackage{orcidlink}
\usepackage[english=american]{csquotes} % choose a suitable language option
\setcitestyle{authoryear,round}

\hypersetup{
	colorlinks=true,
	linkcolor=red,
	linkbordercolor=black,
	citecolor=blue,
	filecolor=magenta,      
	urlcolor=magenta,
	pdfpagemode=FullScreen,
}

\newcommand{\Mycite}[1]{\citeauthor{#1}~(\citeyear{#1})}
\newcommand{\Mycitep}[1]{\citeauthor{#1}~\citeyear{#1}}
\newcommand{\beq}{\begin{equation}}
\newcommand{\eeq}{\end{equation}}
\newcommand{\beqn}{\begin{eqnarray}}
\newcommand{\eeqn}{\end{eqnarray}}

\def \apj {ApJ}

\def \aap {A\&A}

\def \mnras {MNRAS}

\def \prd {Phys. Rev. D}

			{\end{center}\end{minipage}\end{flushleft}} 
		

\begin{document}
%%%%%%%%%%%%%%%%%%%%%%%%%%%%%%%%%%%%%%%%%%%%%%%%%%%%%%%%%%%%%%%%%%%%%%%	
\title{The Guided Moments formalism: a new efficient full-neutrino treatment for astrophysical simulations}
%%%%%%%%%%%%%%%%%%%%%%%%%%%%%%%%%%%%%%%%%%%%%%%%%%%%%%%%%%%%%%%%%%%%%%%	
\author{ M.~R.~Izquierdo$^{1,2}$\orcidlink{0000-0001-9225-3141} \footnote[1]{Corresponding author:\href{mailto:manuel.rodriguez@uib.es}{manuel.rodriguez@uib.es}}, F.~Abalos$^{1,2}$\orcidlink{0000-0001-7863-3711} and C. Palenzuela$^{1,2}$\orcidlink{0000-0003-0634-531X}}
	\affiliation{${^1}$Departament  de  F\'{\i}sica,  Universitat  de  les  Illes  Balears,  Palma  de  Mallorca, E-07122,  Spain}
	\affiliation{$^2$Institute of Applied Computing \& Community Code (IAC3),  Universitat  de  les  Illes  Balears,  Palma  de  Mallorca, E-07122,  Spain}
%%%%%%%%%%%%%%%%%%%%%%%%%%%%%%%%%%%%%%%%%%%%%%%%%%%%%%%%%%%%%%%%%%%%%%%
\begin{abstract}
We present the new Guided Moments (\texttt{GM}) formalism for neutrino modeling in astrophysical scenarios like core-collapse supernovae and neutron star mergers. The truncated moments approximation (\texttt{M1}) and Monte-Carlo (\texttt{MC}) schemes have been proven to be robust and accurate in solving the Boltzmann's equation for neutrino transport. However, it is well-known that each method exhibits specific strengths and weaknesses in various physical scenarios. The \texttt{GM} formalism effectively solves these problems, providing a comprehensive scheme capable of accurately capturing the optically thick limit through the exact \texttt{M1} closure and the optically thin limit through a \texttt{MC} based approach. In addition, the \texttt{GM} method also approximates the neutrino distribution function with a reasonable computational cost, which is crucial for the correct estimation of the different neutrino-fluid interactions. Our work provides a comprehensive discussion of the formulation and application of the \texttt{GM} method, concluding with a thorough comparison across several test problems involving the three schemes (\texttt{M1}, \texttt{MC}, \texttt{GM}) under consideration.
\end{abstract} 
%%%%%%%%%%%%%%%%%%%%%%%%%%%%%%%%%%%%%%%%%%%%%%%%%%%%%%%%%%%%%%%%%%%%%%%
%
\maketitle
%	
%\footnote[1]{Corresponding author:\href{mailto:manuel.rodriguez@uib.es}{manuel.rodriguez@uib.es}}
%
%\small{{\bf Key words:} software: simulations – methods: numerical – stars: neutron – neutrinos – radiative transfer – relativistic processes.}
\normalsize
%
%%%%%%%%%%%%%%%%%%%%%%%%%%%%%%%%%%%%%%%%%%%%%%%%%%%%%%%%%%%%%%%%%%%%%%%
\section{Introduction}
The coalescence of neutron stars is an extraordinary phenomenon in high-energy astrophysics, occurring in extreme environments marked by strong self-gravity, high densities, and elevated temperatures. The simultaneous detection of gravitational waves (GWs) and electromagnetic (EM) signals from binary neutron stars (BNS), exemplified by the groundbreaking GW170817 event (\Mycitep{2041-8205-848-2-L12}; \Mycitep{PhysRevLett.119.161101}), shows the potential of multi-messenger astronomy. Numerous detections of GWs and EM signals from neutron star mergers are expected in the coming years (see, for instance, \Mycitep{Colombo_2022}). The observation of such signals is predicted to have a profound impact on both astrophysics and fundamental physics. Nevertheless, the complexity of modeling all the involved physical interactions constrains our capacity to realistically simulate these binary mergers and compare them with present observations. Unraveling the physics embedded in these signals requires solving, at least approximately, the general relativistic radiation magneto-hydrodynamic equations: Einstein equations for depicting strong gravity, relativistic magnetohydrodynamics (MHD) to model magnetized fluids, and Boltzmann's equation to describe the production and transport of neutrinos. Obtaining accurate solutions for these equations, in realistic general astrophysical settings like neutron star mergers, is an exceedingly challenging task which can solely be addressed through numerical simulations.

After the merger, a massive neutron star or a black hole, surrounded by a strongly magnetized hot and dense accretion torus is likely to be formed (see, for instance, \Mycitep{PhysRevD.97.124039}; \Mycitep{foucart_review}). Throughout the coalescence, various regions may experience conditions ranging from the {\emph{optically thin regime}}, characterized by freely streaming neutrinos, to the {\emph{optically thick regime}}, where neutrinos are mostly trapped and propagate through diffusion. Due to their substantial energies and luminosities, neutrinos are expected to play a fundamental role in  physical processes in the post-merger phase of a neutron star merger (for an extensive review, see \Mycitep{foucart_review}). In particular, neutrino interactions are expected to induce changes in matter composition, influencing the conditions relevant to the $r$-process nucleosynthesis (e.g., \Mycitep{Lippuner_2015}; \Mycitep{Just_2015}; \Mycitep{Thielemann_2017}; \Mycitep{Perego_2021}). Neutrinos produced in  hot and dense matter regions will diffuse and eventually decouple from the fluid at lower densities, being emitted from the system while carrying away energy. As neutrinos extract energy from the system, they could give rise to additional matter outflows, manifested as neutrino-driven winds (e.g., \Mycitep{Dessart_2008}; \Mycitep{Perego_2014}; \Mycitep{Fujibayashi_2017}; \Mycitep{Fujibayashi_2020}).

Accurately capturing all these phenomena in neutron star merger simulations require an accurate and realistic treatment of neutrino transport in numerical relativity (NR) codes (\Mycitep{foucart2022snowmass2021}). This task involves the tremendous challenge of solving the 7-dimensional Boltzmann's equation, which describes the evolution of a distribution function for each neutrino species. To address this complexity, several approaches, including direct and approximate methods have been explored. Presently, various methods directly attempt to solve the full Boltzmann's equation in full GR, using Monte-Carlo (\texttt{MC}) based approaches (see, for example, \Mycitep{Abdikamalov_2012}; \Mycitep{Richers_2015}; \Mycitep{ryan2015bhlight}; \Mycitep{fou_moment}; \Mycitep{Foucart_2018v2}; \Mycitep{Ryan_2020}; \Mycitep{Foucart_2021}; \Mycitep{kawaguchi22}; \Mycitep{fou22}), lattice-Boltzmann methods (\Mycitep{weihv2}), expansion of momentum-space distributions into spherical harmonics ($P_n$) methods (see, e.g., \Mycitep{pomraning1973}; \Mycitep{mcclarren2010robust}; \Mycitep{RADICE2013648}), discrete-ordinates ($S_n$) methods (see, e.g., \Mycitep{pomraning1969}; \Mycitep{mihalas1999}; \Mycitep{Chan_2020}) and a finite element approach in angle (\Mycitep{BHATTACHARYYA2023112365}).
Direct methods, although expected to converge to the true solution, often present challenges in their implementation and computational cost,  making them in most of the cases impractical for providing a numerical solution with sufficient accuracy. A notable example is the handling of the optically thick regime in the Monte-Carlo based methods, where resolving the small neutrino mean-free path becomes challenging, often requiring further simplifications (see, for instance, \Mycitep{fleck_1971}; \Mycitep{fleck_1984}; \Mycitep{wollaber2008}; \Mycitep{Richers_2015}; \Mycitep{Richers_2017};  \Mycitep{fou_moment}; \Mycitep{Foucart_2021}; \Mycitep{foucart_review}). 

On the other hand, approximate methods strike a balance between accuracy and computational efficiency. Neutrino leakage schemes (see, for example, \Mycitep{ruffert96}; \Mycitep{rosswog03}; \Mycitep{Sekiguchi_2010}; \Mycitep{palenzuela_2014}; \Mycitep{palenzuela_2015}; \Mycitep{Perego_2016}; \Mycitep{most2019}; \Mycitep{leakage22}) are a widely used approach due to their computational inexpensiveness. However, the absence of neutrino re-absorption may lead, in certain scenarios, to crude estimations in the amount and composition of the ejecta. A more sophisticated approximation, known as the truncated moment formalism (\Mycitep{thorne81}; \Mycitep{shi11}),  involves the evolution of the lowest moments of the neutrino distribution function (see, e.g.,  \Mycitep{sado13}; \Mycitep{mck14}; \Mycitep{Wanajo_2014}; \Mycitep{fou15}; \Mycitep{just15}; \Mycitep{Sekiguchi_2015}; \Mycitep{kuroda16}; \Mycitep{Foucart_2016}; \Mycitep{Skinner_2019}; \Mycitep{Melon_2019}; \Mycitep{weih20}; \Mycitep{rad2022}; \Mycitep{paper_m1}; \Mycitep{cheong2023}; \Mycitep{musolino2023}; \Mycitep{Schianchi2023}). Typically, only the first two moments are evolved, such that it is often referred to as the \texttt{M1} scheme. This approach requires an algebraic closure for computing the higher moments, which is only known in some limit cases. Additionally, the evolution system can be simplified significantly by deriving evolution equations for the energy-integrated moments, transforming the 7-dimensional Boltzmann's equation into a $3 + 1$ system which resembles the hydrodynamic equations. Unfortunately, even in this simplified scenario, numerical and mathematical challenges persist. Specifically, the truncated moment equations may contain potentially stiff source terms arising from neutrino-matter interactions, causing the equations to shift from hyperbolic to parabolic type in optically thick regions. While this issue can be addressed by employing an Implicit-Explicit Runge-Kutta (IMEX) time integrator (\Mycitep{ascher97}; \Mycitep{pareschi00}; \Mycitep{kennedy03}; \Mycitep{pareschi05}), achieving a balance between accuracy and stability remains a critical consideration (for more details about IMEX schemes within the \texttt{M1} formalism we refer to \Mycitep{weih20} and \Mycitep{paper_m1}). Additionally, moment-based schemes may produce unphysical shocks in regions where radiation beams intersect, leading to solutions that differ from the true solutions of Boltzmann's equation (see for instance the discussion in \Mycitep{foucart_review}).

In this work, we introduce the \emph{\textbf{Guided Moments}} (\texttt{GM})  formalism, inspired in the seminal work presented in \Mycite{fou_moment} and the \emph{Moment Guided Monte-Carlo method} developed in (\Mycitep{degond2011}; \Mycitep{dimarco2013}).  Our approach exploits the benefits of the approximate truncated moment formalism, incorporating evolution equations similar to those in hydrodynamics and ensuring accuracy in the optically thick regime. Additionally, it incorporates aspects of the Monte-Carlo method, providing convergence to the exact solution and accuracy in the optically thin regimes. By exploiting the advantages of each method, we overcome their individual limitations. In this context, we combine the \texttt{M1} and \texttt{MC} approaches to enhance the overall efficacy of our formalism. The fundamental idea is to close the \texttt{M1} evolution equations by computing the second moment with information from the \texttt{MC} solution, especially in the optically thin regime where the \texttt{M1} closure is not exact. In contrast, an exact analytical closure for the second moment exist in the optically thick limit, such that the \texttt{M1} formalism can provide a more accurate and cost-effective solution than the one obtained from the \texttt{MC} scheme, which suffers in this regime. A suitable projection of the \texttt{MC} distribution function, such that the \texttt{MC} moments match the evolved \texttt{M1} ones, is enough to ensure convergence to the exact solution and mitigate the statistical noise so characteristic of the \texttt{MC}-based methods, among other advantages. The resulting scheme outperforms both the \texttt{M1} and \texttt{MC} approaches, providing a comprehensive and accurate solution in both regimes. 

The remainder of the work is organized as follows. In Sec.~\ref{sec2}, we provide an overview of the two formalisms (\texttt{M1} and \texttt{MC}) essential for constructing the guided moments (\texttt{GM}) formalism. Following that, Sec.~\ref{sec3} is dedicated to the presentation and discussion of the complete \texttt{GM} formalism. In Sec.~\ref{sec4}, we present several test problems designed to validate both the validity and accuracy of our formalism. Finally, we present our conclusions and describe possible improvements on the \texttt{GM} formalism. Additionally, in Appendix~\ref{N_app}, we include the additional steps to incorporate also the neutrino number density in our formalism. Appendix~\ref{tetradMC}
describes the tetrad transformation necessary to set the 4-momentum of the neutrinos in the \texttt{MC} scheme.%
%\vspace{-0.8em}
%
%%%%%%%%%%%%%%%%%%%%%%%%%%%%%%%%%%%%%%%%%%%%%%%%%%%%%%%%%%%%%%%%%%%%%%%
\section{The building blocks: truncated moments (M1) and  Monte-Carlo (MC)}\label{sec2}
The neutrino radiation transport can be fully described by the distribution function of neutrinos $f_{\nu}(x^a, p^a)$ in the 7-dimensional phase space given by the spacetime coordinates $x^a = (t, x^i )$ and the neutrino 4-momentum $p^a$ satisfying approximately the null condition $p^a p_a= 0$. This distribution function allow us to define the number of neutrinos within a 6D volume of phase space\footnote{Notice that the momentum volume $\frac{dp^3}{h^3 \sqrt{-g} p^t}$ and the spatial volume $dx^3 \sqrt{-g} p^t$ are invariant under coordinate transformations, such that the 6D volume in the phase space $dx^3 dp^3$ is also invariant.}
\begin{equation}
	N(t) = \int dx^3 \frac{dp^3}{h^3} f_{\nu} (t, x^i, p_j)~.
\end{equation}
One can also define the neutrino stress-energy tensor
\begin{equation}\label{stress_energy_nu}
	T^{a b}(t,x^i) = \int \frac{dp^3}{h^3 \sqrt{-g} p^t} p^a p^b f_{\nu} (t, x^i, p_j)~,
\end{equation}
where $g$ is the determinant of the spacetime metric $g_{ab}$.  
This radiation stress-energy tensor can also be interpreted as the energy-integrated second moment  of the distribution function $f_{\nu}$, as  defined in (\Mycitep{thorne81}; \Mycitep{shi11})\footnote{Roughly speaking, $k$-order moment definition includes the integral of $k$ moments $p^{a_1}...p^{a_k}$, multiplied by   the weighted distribution function $f_{\nu}/\nu^{k-2}$ and using the invariant integration volume in momentum
space $\frac{dp^3}{h^3 \sqrt{-g} p^t}=\nu d\nu d\Omega$, where $\nu$ is the neutrino energy.
The energy-dependent moments are obtained by integrating only the solid angle $d\Omega$ on an unit sphere. When this integration includes also $d\nu$, they are called $k$-order energy-integrated moments (\Mycitep{thorne81}; \Mycitep{foucart_review}).}. In addition, in these works was shown that any given moment includes the lower rank ones, meaning that the (energy-integrated) zeroth and first  moments can be constructed from projections of this stress-energy tensor.

The distribution function for each neutrino species\footnote{By neutrino species we refer to electron, muon and tau neutrinos $(\nu_e, \nu_{\mu}, \nu_{\tau})$ and their respective antineutrinos $({\bar \nu_e}, {\bar \nu_{\mu}}, {\bar \nu_{\tau}})$.} evolves according to Boltzmann's equation
\begin{equation}\label{Boltzmann}
	p^a \left[ \frac{\partial f_{\nu}}{\partial x^a}  - \Gamma^b_{ac} p^c \frac{\partial f_{\nu}}{\partial p^b} \right] = 
	\left[ \frac{\partial f_{\nu}}{\partial \tau}\right]_{coll}~,
\end{equation}
where the right-hand side includes all collisional processes (i.e., emission, absorption and scattering) and $\Gamma^a_{bc}$ are the Christoffel symbols associated to the space-time metric $g_{ab}$. 

Solving the Boltzmann's equation requires, therefore, the time evolution of a 6-dimensional function, a very demanding computational challenge even with modern facilities. Here we will consider two of the different approximations for solving efficiently the Boltzmann's equation: the truncated moment formalism (\texttt{M1}), in which only the lowest moments of the distribution function are evolved, and Direct Simulation Monte-Carlo methods (\texttt{MC}) that attempt to solve directly the Boltzmann's equation. We first give a short summary of these formalisms, and continue with a new method that profits from combining efficiently both of them.

In what follows we will use the standard $3+1$ decomposition to write the covariant equations explicitly as an evolution system of partial differential equations. First, the spacetime metric is decomposed as
\begin{equation}
	ds^2 = -\alpha^2\,dt^2 
	+ \gamma_{ij}\left(dx^i + \beta^i\,dt\right)\left(dx^j + \beta^j\,dt\right)\,,
\end{equation}
where $\alpha$ is the lapse function, $\beta^{i}$ the shift vector, $\gamma_{ij}$ the induced 3-metric on each spatial slice, and $\gamma$  its determinant. Within this decomposition, the normal vector to the hypersurfaces is just $n_a = (-\alpha, 0)$. We also use the standard definition of the extrinsic curvature $K_{ij}$ as the Lie derivative of $\gamma_{ij}$ along the normal vector $n^a$. Throughout this paper, from now on, we will adopt units $G=c=h=1$.
%
%%%%%%%%%%%%%%%%%%%%%%%%%%%%%%%%%%%%%%%%%%%%%%%%%%%%%%%%%%%%%%%%%%%%%%%
\subsection{Truncated Moment Formalism: M1 approach}
%%%%%%%%%%%%%%%%%%%%%%%%%%%%%%%%%%%%%%%%%%%%%%%%%%%%%%%%%%%%%%%%%%%%%%%
%
The truncated moment formalism evolves the lowest moments of the neutrino distribution function. In particular, the \texttt{M1} scheme considers the evolution of the first two moments of the distribution function, which still depend on the neutrino energy (i.e., but not on the direction of the neutrino 4-momentum). While the moment formalism can theoretically accommodate a discretization in neutrino energies, this introduces an additional dimension that significantly escalates the required computational resources. To reduce this computational cost and avoid further technical complexities, it is common to adopt the \emph{grey} approximation, where we primarily focus on evolving energy-integrated moments\footnote{Notice that the neutrino energy spectra is required to compute the energy-averaged opacities/emissivities. Since this energy spectra is only known exactly in the optically thick regime,  an ``energy'' closure is necessary to complete the \texttt{M1} grey approximation. The choice of such closure might have a significant impact in the electron fraction of the ejecta produced in neutron star mergers.}. A more detailed description of the formalism and our implementation is presented in \Mycite{paper_m1}.
 
Let us define the fields which describe our evolution equations in terms of the neutrino radiation stress-energy tensor, which as mentioned before contains also the lower-order (energy-integrated) moments. The interaction between neutrino and matter is usually simplified in the fluid rest frame. Therefore, a convenient decomposition of this tensor in terms of the spatial projection of the second and lower moments is given by
\begin{equation}
	T^{ab} = J u^a u^b + H^a u^b + H^b u^a + Q^{ab}\,,
\end{equation}
where the energy density $J$, the flux density $H^a$ and the symmetric pressure tensor $Q^{ab}$ of the neutrino radiation are computed by an observer co-moving with the fluid. Notice that, by construction, both the flux and the pressure tensors are orthogonal to the fluid velocity, i.e., $H^a u_a = Q^{ab} u_b = 0$. 

In order to obtain a well-posed system of partial differential equations in conservative form is preferable to decompose the same tensor $T^{ab}$ in the inertial frame, namely
\begin{equation}
	T^{ab} = E n^a n^b + F^a n^b + F^b n^a + P^{ab}\,,
\end{equation}
where the radiation energy density $E$ (i.e., the energy-integrated zeroth moment), the radiation flux $F^{a}$ (i.e., the spatial projection of the energy-integrated first moment), and the symmetric radiation pressure tensor $P^{ab}$ (i.e., the spatial projection of the energy-integrated second moment), are evaluated now by normal observers. Note that $F^{a}$ and $P^{a b}$ are orthogonal to $n_a$ by construction, i.e.,  $F^a n_a=0=P^{ab} n_b = 0$. 

We can express the fluid-rest frame quantities $\left\{ J, H^a, Q^{a b}\right\}$ in terms of the Eulerian ones $\left\{ E, F_a, P_{ab}\right\}$  by decomposing the fluid four-velocity as $u^{a} = W(n^{a} + v^{a})$, where $W =- n_{a} u^{a}$ is the Lorentz factor and $v^{a}$ the spatial velocity of the fluid (the full expressions of these frame transformations can be found, for instance, in \Mycitep{paper_m1}; \Mycitep{rad2022}). As usual, the normalization of the four-velocity $u_a u^a =-1$ allow us to compute explicitly the Lorentz factor as $W = (1 - v_i v^i)^{-1/2}$. 
In particular, we will use the relations
\begin{eqnarray}
	J &=& W^2 (E - 2 F^{a} v_{a} + P^{ab} v_{a}v_b)\,,  \\
	H^{a} &=& W( E - F^b v_b)h^{a}_{~c} n^c + W h^{a}_{~b} ( F^b - v_c P^{bc})\,, 
%	Q^{ab} &= T^{cd}_{\rm{rad}}\,h^{a}_{~c}
%	h^b_{~d}\,,
\end{eqnarray}
where $h_{ab} \equiv g_{ab} + u_{a} u_b$ is the projection tensor orthogonal to the four-velocity of the fluid, $h_{ab} u^{a}=0$. 
Notice that when the fluid is at rest, $v_i = 0$, the translation between frames is trivial, i.e., $E=J$, $F^{a} = H^{a}$ and $P^{a b} = Q^{a b}$.

The conservation of energy and linear momentum implies that
\begin{equation}
	\nabla_b T^{ab} = {\cal S}^a\,, \label{eq_conservation}
\end{equation}
where $\nabla$ is the covariant derivative operator compatible with the spacetime metric $g_{a b}$ and ${\cal S}^a$ is the term representing the interaction between the neutrino radiation and the fluid. This term can be written as
\begin{equation}\label{source_term}
	{\cal S}^a = (\eta - \kappa_a J) u^a - (\kappa_a + \kappa_s) H^a\,,
\end{equation}
where $\eta$ is the energy-averaged neutrino emissivity and $(\kappa_a,\kappa_s)$ are the energy-averaged absorption and scattering opacities.

The conservation Eq.~(\ref{eq_conservation}) in the $3+1$ decomposition (\Mycitep{shi11}) can be written as
\begin{eqnarray}
	&\partial_t (\sqrt{\gamma} E)&  \label{eq_evolE}
	+ \partial_i \left[ \sqrt{\gamma} (\alpha F^i - \beta^i E)  \right] = \\
	& & \alpha \sqrt{\gamma} \left[ P^{ij} K_{ij}
	- F^i (\partial_i \alpha)/\alpha - {\cal S}^a n_a  \right]\,, \nonumber \\
	&\partial_t (\sqrt{\gamma} F_i)& \label{eq_evolF}
	+\partial_j \left[ \sqrt{\gamma} (\alpha {P^j}_{i} - \beta^j F_i)  \right] = \\
	& & \sqrt{\gamma} \left[ - E \partial_i \alpha + F_j \partial_i \beta^j + \frac{\alpha}{2} P^{kj} \partial_i \gamma_{kj} + \alpha {\cal S}^a \gamma_{ia}  \right]\,.   \nonumber
\end{eqnarray}	
This system strongly resemble the hydrodynamical equations except by the fluid-neutrino interaction term ${\cal S}^a$, which is given in the $3+1$ decomposition by%
\begin{align}
	{\cal S}_n &= - {\cal S}^a n_a =  W \left[ (\eta + \kappa_s J) - (\kappa_a + \kappa_s) (E - F_i v^i) \right]~, \nonumber \\
	{\cal S}_i &= {\cal S}^a \gamma_{ia} = W (\eta - \kappa_a J) v_i - (\kappa_a + \kappa_s) H_i~.
\end{align}
The evolution equations Eqs.~($\ref{eq_evolE}$, $\ref{eq_evolF}$) represent the \texttt{M1} formalism. Notice that they are exact, although with two limitations: (i) there is no closed form due to the unknown second moment $P^{ij}$ (or conversely $Q^{ij}$), and (ii) the energy-averaged emissivities/opacities $(\eta,\kappa_a,\kappa_s)$ require information about the energy spectra which is not computed explicitly within the formalism.
%
%%%%%%%%%%%%%%%%%%%%%%%%%%%%%55
\subsubsection{Closure Relations}
In general, the second moment $P^{ij}$ depends not only on the local values of the lower moments $(E,F^k)$ at a point, but also on the global geometry of these fields in a neighborhood of that point. Although  approximate closures can be defined at different limits, there is no single closure prescription $P^{ij}=P^{ij}(E,F^k)$ that describes accurately all possible regimes. Instead, the \texttt{M1} scheme usually employs an analytic closure that interpolates $P^{ij}$ between two limits: the \emph{optically thick limit} (where matter and radiation are in thermodynamic equilibrium) and the \emph{optically thin limit} (where propagation of radiation, in a transparent medium, comes from a single point source). Fortunately, there are explicit expressions for $P_{ij}$ in both limits, $\left\{P_{ij}^\text{thin}, P_{ij}^\text{thick} \right\}$ that give the following closure 	
\begin{widetext}
\begin{equation}\label{eq_Ptot}
	P^{\texttt{M1}}_{ij} = \frac{3 \chi(\xi) - 1}{2} P^{{\text{thin}}}_{ij} + \frac{3 \left[1- \chi(\xi)\right]}{2} P^{{\text{thick}}}_{ij} = d_{\text{thin}}P^{{\text{thin}}}_{ij} + d_{\text{thick}}P^{{\text{thick}}}_{ij} \,,
\end{equation}
\end{widetext}
where $\chi(\xi) \in \left[\frac{1}{3}, 1\right]$ is the variable Eddington factor and $\xi \in [0,1]$ the  norm of the normalized flux (for more details see \Mycitep{levermore81}).

As discussed in \Mycite{shi11}, there are several relativistic generalizations for $\xi$, but the only one that is accurate in the optically thick limit is
\begin{equation} \label{eq_xi}
	\xi \equiv \sqrt \frac{H_a H^a}{J^2}\,.
\end{equation}
Unfortunately, this choice is computationally expensive, since the calculation of $\xi$ requires a root-finding method to transform the fields from the inertial to the fluid-rest frame.  A detailed root finding routine was used and explained in \Mycite{paper_m1}.

We choose the commonly employed \textit{Minerbo closure} (\Mycitep{minerbo79}), which is exact in both limits,
\begin{equation}\label{eq_minerbo}
	\chi (\xi)= \frac{1}{3} + \xi^2 \left(
	\frac{6 - 2\xi + 6 \xi^2}{15} \right)\,.
\end{equation}
Notice however that there are many other choices (see \Mycitep{mur17})  that might also capture accurately these two limits.

In the optically thick limit (or diffusion limit), the radiation pressure tensor measured by the fluid-rest frame is simply described by $	Q^{{\text{thick}}}_{ab} = J_{{\text{thick}}} h_{ab}/ 3~.$ To obtain the pressure tensor $P^{{\text{thick}}}_{ij}$ in the inertial frame, first we need to compute the intermediate quantities, $\left\{J_{\text{thick}}, H^{i}_{\text{thick}}, H^{n}_{\text{thick}}\right\}$, that relate the fluid rest frame with the inertial frame. The final closure can be written as
\begin{eqnarray}
	&&P_{ij}^{{\text{thick}}} = \frac{4}{3} J_{{\text{thick}}} W^2 v_i v_j \nonumber  \\
	&&~~~~	+ W (H^{{\text{thick}}}_i v_j + H^{{\text{thick}}}_j v_i) + \frac{1}{3} J_{{\text{thick}}} \gamma_{ij} \,,
	\label{eq_Pthick} \\
	&&J_{{\text{thick}}} = \frac{3}{2 W^2 + 1} \left[ (2 W^2 - 1) E - 2 W^2 F^k v_k \right]\,, 
	\nonumber \label{eq_Jthick} \\
	%	&& \gamma^a_b H^b_{\text{thick}} = \frac{F^a}{W} + \frac{W v^a}{2 W^2 +1} \left[ (4 W^2 +1) v_k F^k - 4 W^2 E \right]\,, \nonumber \\
%	&&H^n_{{\text{thick}}} = -H^a n_a = W ( E - J_{{\text{thick}}} - F^k v_k) \,,\\
	&&H^i_{{\text{thick}}} =  \frac{F^i}{W} + \frac{W v^i}{2 W^2 + 1} 
	\left[(4 W^2 + 1) F^k v_k - 4 W^2 E \right]\,. \nonumber 
\end{eqnarray}
In these regions the radiation satisfies $H^a  \approx 0$ (i.e., $\xi \approx 0$ and $\chi \approx 1/3$), so $P^{\texttt{M1}}_{ij} \approx P^{{\text{thick}}}_{ij}$. 

In the optically thin limit, we assume that radiation is streaming at the speed of light in the direction of the radiation flux, leading to the explicit relation
\begin{equation}\label{eq_Pthin}
	P^{{\text{thin}}}_{ij} = \frac{F_i F_j}{F^k F_k} E \,.
\end{equation}
In these regions, $H^a H_a  \approx J^2$ (i.e., $\xi \approx 1$ and $\chi \approx 1$), so $P^{\texttt{M1}}_{ij} \approx P^{{\text{thin}}}_{ij}$. Unfortunately, this limit is not unique, as it is determined by the non-local geometry of the radiation field. In general, $P^{{\text{thin}}}_{ij}$ will not correctly describe free-streaming neutrinos produced by multiple sources, since in vacuum they usually do not propagate in the same direction. For instance, the trajectory of colliding beams will propagate unphysically in the direction of their average momentum, as it is very well represented by the well-known \emph{double beam test} (see, e.g., \Mycitep{sado13}; \Mycitep{fou15}; \Mycitep{weih20}), which will be discussed in Sec.~\ref{T2}. This behavior is responsible of un-physical radiation shocks and lead, in general, to incorrect results in some scenarios (for further details, see \Mycitep{foucart_review}). Dealing with these issues requires going beyond the \texttt{M1} formalism. In Sec.~\ref{T2}, we illustrate how both the \texttt{MC} scheme and our \texttt{GM} formalism sidestep this problem, allowing both beams to follow their original paths without interference.
%
%%%%%%%%%%%%%%%%%%%%%%%%%%%%%%%%%%%%%%%%%%%%%%%%%%%%%%%%%%%%%%%%%%%%%%%
\subsection{Direct Simulation Monte-Carlo: MC approach}
Monte-Carlo (\texttt{MC}) methods attempt to sample the distribution function $f_{(\nu)}$ with a discrete set of \emph{neutrino packets}  describing a large numbers of neutrinos that interact with the matter and propagate through the numerical grid. In a \texttt{MC} simulation, the ensemble of $N_{\text{T}}$  packets at time $t$, each labeled by the index $k$ and containing $N_k$ neutrinos located at spatial coordinates $x^i_k$ with 4-momentum $p^a_k$, serves as an approximation of the distribution function\footnote{It is important to highlight that using lower indices in $p_i$ and upper indices in $x^i$ is essential to maintain the relativistic invariance of this equation. 
}
\begin{equation}\label{discrete_PDF}
	f_{\nu} \sim f_{(\nu)} = \sum_{k=1}^{N_\text{T}} N_k \delta^3(x^i-x^i_k) \delta^3(p_i-p_i^k)~.
\end{equation}

Given this sampled distribution function, we can compute the radiation stress-energy tensor averaged in the neighborhood of the point $x^i$ (i.e., for instance a grid cell of volume $\Delta V = \Delta x^3$), namely
\begin{eqnarray}
	{\bar T}^{ab} (t, x^i) = \sum_{k \in \Delta V} N_k \frac{p_k^a p_k^b}{\sqrt{-g} \Delta V p_k^t}  ~,
\end{eqnarray}
where the summation only involves neutrino packets located within the volume $\Delta V$. Notice that the \texttt{MC} pressure tensor reduces just to
\begin{eqnarray}
	P^{\texttt{MC}}_{ij} = {\bar P}_{ij} &=& \sum_{k \in \Delta V} N_k \frac{p_i^k p_j^k}{\sqrt{-g} \Delta V p_k^t}~.
\end{eqnarray}

In a Monte-Carlo transport scheme, Boltzmann's equation for $f_{(\nu)}$ can be translated into prescriptions for the creation, annihilation, scattering and propagation of the neutrino packets sampling $f_{(\nu)}$.  These prescriptions are explained in more detail in the following subsections. 
%
%%%%%%%%%%%%%%%%%%%%%%%%%%%%%%%%%%%%%%%%%%
\subsubsection{Free propagation}
%%%%%%%%%%%%%%%%%%%%%%%%%%%%%%%%%%%%%%%%%%
%
In the absence of collisional terms, a direct substitution of the 
discrete distribution function Eq.~(\ref{discrete_PDF}) into the Boltzmann's equation Eq.~(\ref{Boltzmann}) implies that neutrino packets follow geodesic equations. Specifically,
\begin{align}
	\frac{d x^\mu}{d\lambda}=p^\mu,\,\,\,
	\frac{d p_i}{d\lambda}=\Gamma^\mu_{~i\nu}p_\mu p^\nu,
\end{align}
with the normalization condition
\begin{align}
	p^a p_a = -m_{\nu}^2 ~,
\end{align}
where $\lambda$ is an affine parameter and $m_{\nu}$ the neutrino mass. 
By using the $3+1$ decomposition, we rewrite these equations in the grid frame in a form suitable for time evolution, 
\begin{align}
	\frac{dx^i}{dt}&=\gamma^{ij}\frac{p_j}{p^t}-\beta^i~,\label{eq:geo1}\\
	\frac{dp_i}{dt}&=-\alpha\left(\partial_i\alpha\right)p^t+\left(\partial_i \beta^j\right)p_j-\frac{1}{2p^t}\left(\partial_i \gamma^{jk}\right)p_jp_k~,\label{eq:geo2}\\
	p^t&=\frac{1}{\alpha}\sqrt{ m_{\nu}^2 +  \gamma^{ij}p_ip_j}~.\label{eq:geo3}
\end{align}
Since the neutrino mass is negligible compared to the energy scale of the system, it is common to assume $m_{\nu}=0$, which implies that the neutrino 4-momentum $p^a$ is null. 
%
%%%%%%%%%%%%%%%%%%%%%%%%%%%%%%%%%%%%%%%%%%
\subsubsection{Emission}
%%%%%%%%%%%%%%%%%%%%%%%%%%%%%%%%%%%%%%%%%%
%
Copious amounts of neutrinos are produced in hot and dense matter. These neutrinos are emitted isotropically in the fluid frame, a process which involves the creation of new neutrino packets in the \texttt{MC} formalism.

For a given neutrino species, if $\eta$ denotes the total emissivity (i.e., the energy of neutrinos emitted per time interval $\Delta t$ and unit volume $\Delta V$), the total energy of the emitted neutrinos is given by
\begin{equation}
	E_{\text{ems}} = \sqrt{-g} \, \Delta V \Delta t \, \eta ~,
\end{equation}
where the emissivity is assumed to remain constant throughout a time step. Following (\Mycitep{fou_moment}; \Mycitep{Foucart_2021};  \Mycitep{kawaguchi22}), the number of neutrino packets emitted in the fluid frame within a given energy bin $[E_{b-1},E_b]$ for a specific value of $\eta_b$ can be approximated as
\begin{equation}\label{Np_def}
	N^{(b)}_p \approx \frac{E_{\text{ems}}}{E_{\text{packet}}} \approx \sqrt{-g} \Delta V \Delta t \frac{\eta_b}{E_{\text{packet}}}~.
\end{equation}%
Since the total emissivity is the sum of the emissivities at each energy bin (i.e., $\eta = \sum_{b} \eta_b$), then the total number of created packets within a cell is just $N_p = \sum_{b} N^{(b)}_p$.
All packets are initialized with the fluid rest frame energy $\nu = (E_{b-1} + E_b)/2$ and represent a number of neutrinos  $N_k=E_{\text{packet}}/\nu$. 

In practice, the procedure for packet creation can be described as follows:
\begin{enumerate}
	\item Compute the energy in thermal equilibrium $E_{\text{thermal}}$ by using the black-body function for the neutrino density $B_{\nu}$.
	In many cases we can use Kirchoff's law $\eta = B_{\nu} \kappa_a$, that is,
	\begin{equation}
		E_{\text{thermal}} = B_{\nu} \Delta V =\Delta V \frac{\eta}{\kappa_a}~,
	\end{equation}
	where we remind that $\kappa_a$ is the absorption opacity.
	
	\item Compute the energy of the neutrino packet in terms of a free parameter $N_{\text{tr}}$, which sets the number of packets required to describe the neutrinos in thermal equilibrium with matter, namely
	\begin{equation}\label{Ntr}
		E_{\text{packet}} = E_{\text{thermal}} / N_{\text{tr}}~.
	\end{equation}
	 Notice that the choice of $N_{\text{tr}}$, will influence how computational resources are distributed on our grid, i.e., to determine whether more or fewer \texttt{MC} packets are generated.

	\item Estimate the number of packets that should be created using the estimate Eq.~(\ref{Np_def}). For a single energy bin and using Kirchoff's law, this relation can be simplified to
	\begin{equation}
		N_p = \frac{E_{\text{ems}}}{E_{\text{packet}}}
		\approx \Delta t \, \kappa_a N_{\text{tr}}~.
	\end{equation}
\end{enumerate}

Once the number of packets has been determined, their setup proceeds as follows: the packet's location is randomly drawn from a homogeneous distribution within the designated cell in the spatial coordinates of the simulation $x^i$. The emission time of neutrinos is randomly selected from a uniform distribution in time, and the emitted neutrinos are subsequently evolved until the end of the current time step. The new neutrino packets are emitted isotropically in the fluid rest frame, such that the neutrino 4-momentum in this frame can be initialized as
\begin{equation}\label{packet_creation}
	p^{a'}_k = \nu_k (1, \sin\theta \cos \phi, \sin\theta \sin \phi, \cos \theta)~.
\end{equation}
We draw $\cos \theta$ from a uniform distribution in $[-1, 1]$ and $\phi$ from a uniform distribution in $[0, 2 \pi]$. 
In order to convert the neutrino 4-momentum from the fluid rest frame to the inertial frame we use the transformations described in the Appendix \ref{tetradMC}.  
%
%%%%%%%%%%%%%%%%%%%%%%%%%%%%%%%%%%%%%%%%%%
\subsubsection{Absorption and elastic scattering}
%%%%%%%%%%%%%%%%%%%%%%%%%%%%%%%%%%%%%%%%%%

The interaction of neutrinos with matter might lead to the absorption of the neutrinos, increasing the energy of the fluid, or to a scattering where the energy is conserved but the propagation direction changes. In the \texttt{MC} formalism, the absorption process involves the destruction of neutrino packets, whereas scattering involves a change in the 4-momentum of the neutrinos.

To evolve a neutrino packet over a time interval $\Delta t_p$, we first determine whether the packet is free-streaming or if it undergoes absorption or scattering by the fluid. The probabilities of absorption and scattering can be computed from the infinitesimal optical depth along a geodesic given by $d \tau = \kappa \nu d\lambda = (\kappa \nu/p^t ) dt$, where $d \lambda$ represents the increment in the affine parameter ($p^a = dx^a/d \lambda$).
The time intervals before the first absorption/scattering are then given by a Poisson distribution (see, e.g., \Mycitep{wollaber2008}),
\begin{equation}
\Delta t_a = - \kappa_a^{-1} \frac{p^t}{\nu} \ln r_a~,~~~
\Delta t_s = - \kappa_s^{-1} \frac{p^t}{\nu} \ln r_s~,
\end{equation}
with $r_s$ and $r_a$ drawn from a uniform distribution in $(\delta, 1]$, being
$\delta$ a very small number. 

We then identify the smallest among the three time intervals ($\Delta t_p, \Delta t_a, \Delta t_s$). If $\Delta t_p$ is the smallest, the packet is propagated by $\Delta t_p$ following the geodesic equations without interacting with the fluid. If $\Delta t_a$ is the smallest, the packet is propagated by $\Delta t_a$ and subsequently absorbed, leading to its removal from the simulation. Lastly, if $\Delta t_s$ is the smallest interval, the packet is propagated by $\Delta t_s$ and then scattered by the fluid. Following scattering events, we begin a new time step with $\Delta t_p \rightarrow \Delta t_p - \Delta t_s$.

Scattering is performed in the fluid rest frame. As we only consider isotropic elastic scattering, we simply redraw the 4-momentum $p^{a'}_k$ (at constant fluid rest frame energy $\nu_k$), from the same isotropic distribution as during the packet creation (i.e., see Eq.~(\ref{packet_creation})). 

%%%%%%%%%%%%%%%%%%%%%%%%%%%%%%%%%%%%%%%%%%%%%%%%%%%%%%%%%%%%%%%%%%%%%%%
\section{Guided Moments Formalism (GM)} \label{sec3}

Our goal is to combine the \texttt{M1} and \texttt{MC} formalisms into a single one, capitalizing on the strengths of each while mitigating their respective weaknesses. 
On one hand, the \texttt{MC} formalism is well known for its remarkable simplicity, since the neutrino interactions can be modeled either by solving ordinary differential equations or relying on simple stochastic processes. Moreover, the error does not scale with the dimensionality of the problem, allowing to achieve in high-dimensional problems (i.e., like Boltzmann's equation) a fairly well resolved solution with a reasonable amount of resources.
%\end{itemize}
However, the \texttt{MC} formalism also has two main drawbacks:
\begin{itemize}
		\item An inherent large statistical error that decays slowly with the total number of packets, ${\cal O}(N_{\text{T}}^{-1/2})$.
		\item A high computational cost for evolving optically thick regions, where many packets are constantly created and reabsorbed.
\end{itemize}

On the other side, the \texttt{M1} formalism (with the grey approximation) stands out for its distinct advantage in evolving equations similar to those found in hydrodynamics, which can be solved accurately with high-order finite difference/volume numerical methods. However, this approach comes with two significant challenges:
\begin{itemize}
	\item The closure for the second moment is known exactly only in two scenarios: either when neutrinos are in thermal equilibrium with the fluid in optically thick regimes, or when there is a single radiation source in the optically thin limit. Beyond these limits, the closure is merely approximated.
	\item Within the grey approximation, emissivities and opacities are averaged over neutrino energies using the Fermi-Dirac distribution function, that is, assuming neutrinos are in local thermal equilibrium with the fluid. This estimate is accurate only in optically thick regimes, but it deviates significantly from the true solution in other regimes.
\end{itemize}

One could combine the strengths of each formalism and avoid their weaknesses by solving them simultaneously.
% and employing the most accurate solution available in the different regimes.
The \texttt{MC} formalism provides the evolution of the neutrino distribution function, that could be employed to compute the closure relation for the second moment as well as the emissivities/opacities in the \texttt{M1} formalism. The \texttt{M1} formalism provides an accurate solution that can be used to reduce the statistical error in the \texttt{MC} solution without increasing significantly the cost of the simulation. In addition, it might also solve the other main issue in \texttt{MC} simulations: the high cost of evolving optically thick regions.
We can simply rely on the \texttt{M1} formalism in optically thick regions, where it is very accurate, and evolve more \texttt{MC} packets only below a certain optical depth. Finally, and most importantly, as the \texttt{MC} evolution asymptotes to the exact solution of Boltzmann’s equation, and since the \texttt{M1} evolution   can take information\footnote{In this context, we mean all the missing information about higher moments and the neutrino energy spectra, that would otherwise be approximated using analytical prescriptions.} from \texttt{MC} solutions, the algorithm for \texttt{M1} converges to a true solution of  Boltzmann's equation for infinite spatial grid resolution and infinite number of neutrino packets.

Actually, the combination of \texttt{M1} and \texttt{MC} formalisms was already attempted in \Mycite{fou_moment}, where all these issues were discussed in detail. However, it was also stated that: {\em it was not clear if the coupled \texttt{MC}-\texttt{M1} system was to be numerically stable, and that any additional work might be required to guarantee that the coupled \texttt{MC}-\texttt{M1} equations are well-behaved for realistic astrophysical simulations}.
We believe that the \texttt{MC}-\texttt{M1} problems that were observed in that work were due to an unbalance in the information flow. The \texttt{M1} equations received information from the \texttt{MC} solution (i.e., by using the discrete distribution function) to compute the closure relation and the energy-averaged emissivities/opacities within the grey approximation. However, there was no feedback from the \texttt{M1} solutions into the \texttt{MC} equations, not even in the optically thick regions where  the former solutions should be fairly accurate. Therefore, it seems natural to think that the missing crucial step consists on passing this information from the \texttt{M1} solutions to the \texttt{MC} distribution function. 

Inspired by the moment guided Monte-Carlo approach (\Mycitep{degond2011}; \Mycitep{dimarco2013}), an advanced numerical method from applied mathematics to combine the kinetic and Euler equations, we have designed a way to project the \texttt{MC} distribution function such that its lowest moments match exactly to the ones evolved by the \texttt{M1} equations. Our {\bf \emph{Guided Moments}} (\texttt{GM}) formalism shares some features with the original method, although the application to relativistic neutrino transport problem presents also many differences. The remaining of this section is devoted to describe the basics of our method and explain how it can be applied specifically for this particular problem of neutrino transport.

%%%%%%%%%%%%%%%%%%%%%%%%%%%%%%%%%%%%%%%%%%%%%
\subsection{Basics of the method}
%%%%%%%%%%%%%%%%%%%%%%%%%%%%%%%%%%%%%%%%%%%5

Our goal is to solve the Boltzmann's equation with a Monte-Carlo formalism, and simultaneously with the truncated moments formalism  by using any type of finite difference or finite volume scheme. We assume that an {\em exact} second moment $P_{ij}$ can be written as the pressure tensor calculated with the \texttt{M1} closure relation $P^{\texttt{M1}}_{ij}$ plus a correction term that can only be evaluated by using the distribution function of the neutrinos. By including this correction, both \texttt{M1} and \texttt{MC} formalisms should provide the same results in terms of macroscopic quantities (i.e., the lowest moments), except for numerical errors. It
is natural to assume that the set of moments ${\cal J}_a$ obtained from the truncated moments formalism represents in general a better statistical estimate than the moments ${\cal {\bar J}}_a$ calculated from the Monte-Carlo method (they will be defined explicitly in Eqs.~(\ref{eq_J_M1}, \ref{eq_J_MC}) below), since the resolution of the truncated moment equations does not involve any stochastic process. 

Let us assume that at time $t^n$ the lowest moments constructed with the sampled \texttt{MC} distribution function $f_{(\nu)}^n$ are consistent with those evolved by the \texttt{M1} formalism (i.e., ${\cal {\bar J}}_a^n = {\cal {J}}_a^n $). Thus, we can summarize the method to evolve the solution from $t^n$ to $t^{n+1}$ in the following way:

\begin{enumerate}
	
	\item Solve the Boltzmann's equation with a \texttt{MC} scheme to obtain an approximated solution $f_{(\nu)}^{*}$ at $t^{n+1}$, which is used for calculating the first set of moments ${\cal {\bar J}}_a^*$. 
	\item Solve the \texttt{M1} equations with a finite volume/difference scheme, using the distribution function $f_{(\nu)}^n$ to calculate the correction term appearing in the closure for the exact second moment. The evolved \texttt{M1} fields allow to reconstruct the second set of moments ${\cal J}_a^{n+1}$ at $t^{n+1}$.
	\item Match the lowest \texttt{MC} moments to the lowest \texttt{M1} ones through a transformation $T$ of the samples values,  $f_{(\nu)}^{n+1}=T(f_{(\nu)}^{*})$, such that ${\cal {\bar J}}_a^{n+1}={\cal J}_a^{n+1}$.\label{step3} 
	\item Restart the computation to the next time step. 
\end{enumerate}
The piece that was missing in previous works was the Step~\ref{step3}. We will focus on that transformation after giving some details on the application of this abstract scheme to the specifics of the neutrino transport problem.

%%%%%%%%%%%%%%%%%%%%%%%%%%%%%%%%%%%%%%%%%%%%%%%%%%%%%%%%%%%%
\subsection{Guided moments for neutrino transport}
%%%%%%%%%%%%%%%%%%%%%%%%%%%%%%%%%%%%%%%%%%%%%%%%%%%%%%%%%%%%

The first step of the guided moments formalism consists just on performing a standard evolution using the \texttt{MC} method, which can be applied directly also to our problem. At the second step, the \texttt{M1} equations are evolved by using the exact $P_{ij}$, which necessarily must included corrections from the sampled distribution function. There is not a unique way to incorporate these corrections, leading to several possible choices. The exact closure which is more similar to the one presented in the original work (\Mycitep{degond2011}; \Mycitep{dimarco2013}) would be
\begin{equation}
P_{ij} = P^{{\text{thick}}}_{ij}  +  \int \frac{dp^3}{h^3 \sqrt{-g} p^t} p_i p_j (f_{\nu} - f^{eq}_{\nu})   ~,
\end{equation}
where $f^{eq}_{\nu}$ is the Fermi-Dirac distribution function for neutrinos in thermodynamical equilibrium with the fluid. Note that to obtain this exact closure we have used
\begin{equation}
P^{{\text{thick}}}_{ij} =  \int \frac{dp^3}{h^3 \sqrt{-g} p^t} p_i p_j f^{eq}_{\nu}   ~,
\end{equation}
although for the evolution one would use the expression given by Eq.~(\ref{eq_Pthick}). Therefore, one would solve $P^{{\text{thick}}}_{ij}$ using finite difference/volume, and the deviations from the equilibrium state with the sampled distribution function.
Another exact closure might be obtained by using information only from the \texttt{MC} formalism, that is, $P_{ij} = P^{{\text{\texttt{MC}}}}_{ij}$. However, this choice would not mitigate the issues related to the  \texttt{MC} scheme.

Although in realistic applications one of the previous options might be more suitable, for simplicity we have chosen a final pressure tensor $P^{\texttt{GM}}_{ij}$ given by the following straightforward combination\footnote{Following \Mycite{fou_moment}, we reduce the statistical error in the \texttt{MC} pressure tensor by modifying it as $P^{\texttt{MC}}_{ij} \rightarrow  (P^{\texttt{MC}}_{ij}/\bar{E}) E$.}
% $P^{\texttt{MC}}_{ij} \rightarrow  (P^{\texttt{MC}}_{ij}/{E}_{\texttt{MC}}) E_{\texttt{M1}}$.}
\begin{equation}\label{Pij_GM}
P^{\texttt{GM}}_{ij} = h(\xi) P^{\texttt{MC}}_{ij} + [ 1-h(\xi) ] P^{\texttt{M1}}_{ij}~,
\end{equation}
being $h(\xi)$ a smooth function of the normalized flux that vanishes in the optically thick limit (i.e., $h(\xi \rightarrow 0) = 0$) and goes to unity in optically thin mediums (i.e., $h(\xi \rightarrow 1) =1$). This splitting allows us not only to recover accurately the optically thick regime with the \texttt{M1} formalism, but also to introduce corrections by using the \texttt{MC} distribution function as the neutrinos move away from that regime.

The third step of our method involves the matching of moments. It is useful to define them as a four dimensional co-vector ${\cal J}_a$ (i.e., equivalent to the first moment, see,  \Mycitep{thorne81}; \Mycitep{shi11}), constructed by projecting the stress-energy tensor with the normal to the hypersurfaces, that is,
\begin{equation}
%	T_{ab} &= En_an_b +F_a n_b + F_b n_a + P_{ab} \\		
 {\cal J}_a^{\texttt{M1}}={\cal J}_a \equiv -T_{ab}n^{b} = E n_a + F_a ~.
\label{eq_J_M1}
\end{equation}	
Clearly, this is a convenient way to deal with the zeroth and the (projected) first moment $(E, F_i)$ within a four-dimensional moment. We denote this quantity as ${\cal J}_a^{\texttt{M1}}$ when it is calculated using the evolved moments from the \texttt{M1} formalism.

We want to match these moments with the equivalent quantities ${\cal {\bar J}}_a$ computed from the \texttt{MC} discrete distribution, namely 
\begin{equation}
{\cal J}_a^{\texttt{MC}} = {\cal {\bar J}}_a \equiv -{\bar T}_{ab}n^{b} = \sum_{k \in \Delta V} N_k \frac{p^k_a}{\sqrt{\gamma}\Delta V}~.
\label{eq_J_MC}
\end{equation}

Using the expression for null neutrino momentum  $p_{a}^{k}=\epsilon_{k}\left(  n_{a}+l_{a}^{k}\right)$, with $n^a l_a^{k} = 0$, the \texttt{MC} moments can be written as
\begin{equation}
{\cal {\bar J}}_a = \sum_{k \in \Delta V}  \frac{{N}_k \epsilon_k}{\sqrt{\gamma}\Delta V}
(n_a + l^k_a)~.
\end{equation}
where $\epsilon_k = -p^k_{a} n^{a}$ is just the neutrino energy measured in the inertial frame.
If we project ${\cal {\bar J}}_a$ along and perpendicular to the normal $n_{a}$, we find that
\begin{equation}
\bar{E} = \sum_{k \in \Delta V}  \frac{{N}_k \epsilon_k}{\sqrt{\gamma}\Delta V}	
~~~,~~~
\bar{F}_{a} = \sum_{k \in \Delta V}  \frac{{N}_k \epsilon_k l^k_a}{\sqrt{\gamma}\Delta V} 
~.\label{eq_F_hat_1}%
\end{equation}
such that the relation  ${\cal {\bar J}}_a=\bar{E} n_{a}+ \bar{F}_{a}$ is recovered, equivalent to that of the \texttt{M1} moments.

We are interested on transforming the neutrino  4-momentum $p_{k}^{a}$ into other 
4-momentum $\tilde{p}_{k}^{a} = \Lambda^a_b \, p_{k}^{b}$ such that
\begin{equation}
{\cal J}_a^{\texttt{M1}} = \sum_{k \in \Delta V} {N}_k \frac{\tilde{p}^k_a}{\sqrt{\gamma}\Delta V}~.
\end{equation}
We can modify $p_{k}^{a}$ to $\tilde{p}_{k}^{a}$ without changing its norm by using a Lorentz boost transformation that converts ${\cal J}_a^{\texttt{MC} }$ into ${\cal J}_a^{\texttt{M1} }$. 
Although this would correspond directly to the original moment guided approach, we have found in our tests that a better option is to choose as a target a combination of the \texttt{M1} and \texttt{MC} moments, namely,
\begin{equation}
{\cal J}^{\texttt{GM}}_a = h(\xi) {\cal J}_a^{\texttt{MC}} + [ 1-h(\xi) ] {\cal J}_a^{\texttt{M1}}~.
\label{eq_J_GM}
\end{equation}
The motivation for this choice is to recover exactly the \texttt{M1} moments in the optically thick limit, while the \texttt{MC} moment remains unchanged in the optically thin limit, providing a better physical model in this regime. An explicit expression for this Lorentz transformation, that sends ${\cal J}_a^{\texttt{MC} }\rightarrow {\cal J}^{\texttt{GM}}_a$, is given in Sec.~\ref{LBL}.

\begin{figure}[t!]%[t!]%[htb!]
	\includegraphics[width=\columnwidth]{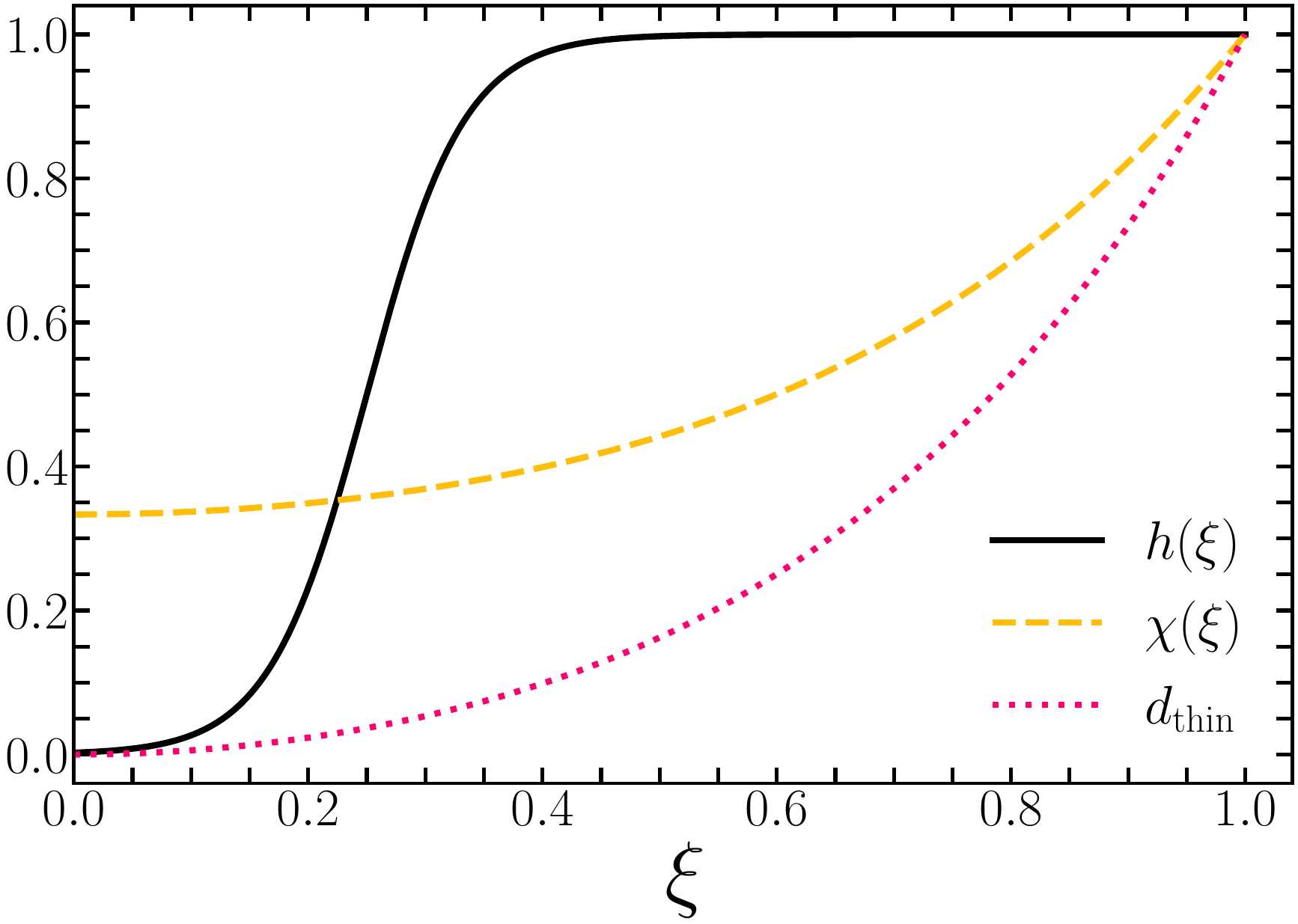}
	\caption{{\em Comparison: $h(\xi)$, $\chi(\xi)$ (Minerbo closure) and $d_{\text{thin}}$. } Logistic function with the parameters set in our tests: a transition at $\xi_0 = 1/3$ with a steepness $k=12$, such that the optically thick regime is fully recovered for $\xi \leq 0.1$ whereas for $\xi \geq 0.5$ the \texttt{MC} pressure tensor fully dominates the solution.
	}
	\label{closure_comparison}
\end{figure}

At this point it is interesting to discuss the different limits of the guided moments formalism, which strongly depends on the profile of the function $h(\xi)$. A suitable choice could be the logistic function (see Fig.~\ref{closure_comparison}), a smooth approximation to the step function with two parameters to set the location of the transition  $\xi_0$ and its steepness $k$, that is,
\begin{equation}\label{h_xi}
h(\xi) = \frac{1 }{1 + e^{-2\,k (\xi - \xi_0)} } ~.
\end{equation}

When $\xi \rightarrow 0$, the pressure tensor is dominated by the one in the optically thick limit (i.e., $P_{ij}^{\texttt{GM}} \rightarrow P_{ij}^{\texttt{M1}} \rightarrow P_{ij}^{\text{thick}}$), while that the projection brings the moments of the \texttt{MC} to match exactly the ones of the \texttt{M1} (i.e., ${\cal J}_a^{\texttt{MC}} \rightarrow {\cal J}_a^{\texttt{GM}}$ and $ {\cal J}_a^{\texttt{GM}} \rightarrow {\cal J}_a^{\texttt{M1}}$). On the opposite limit $\xi \rightarrow 1$, the pressure tensor is dominated by the \texttt{MC} one (i.e., $P_{ij}^{\texttt{GM}} \rightarrow P_{ij}^{\texttt{MC}}$) since in optically thin regimes the \texttt{M1} solution is only valid for isolated distant sources. In this limit only the \texttt{MC} solution converges to the exact solution. This is consistent with the projection on the moments, reducing just to the identity in this limit (i.e., ${\cal J}_a^{\texttt{MC}} \rightarrow {\cal J}_a^{\texttt{GM}}$ and ${\cal J}_a^{\texttt{GM}} \rightarrow {\cal J}_a^{\texttt{MC}}$).

%%%%%%%%%%%%%%%%%%%%%%%%%%%%%%%%%%%%%%%%%%%%%%%%%%%%%%%%%%%%%%%%%%%%%%%
\subsection{Matching the Moments: Lorentz Boost Link} \label{LBL}
%%%%%%%%%%%%%%%%%%%%%%%%%%%%%%%%%%%%%%%

We can modify $p_{k}^{a}$ to $\tilde{p}_{k}^{a}$ without changing its norm by using a Lorentz boost transformation that sends ${\cal {\bar J}}_a={\cal J}_a^{\texttt{MC}}$ into ${\cal J}_a={\cal J}_a^{\texttt{GM}}$. Since Lorentz transformations (defined as $\Lambda_{b}^{a} g^{bc} \Lambda_{c}^{d} =g^{ad}$) preserve vector norms, and ${\cal J}_a^{\texttt{MC}}$ is in general timelike and future-directed, for the desired transformation to exist, ${\cal J}_a^{\texttt{GM}}$ has to be timelike and future-directed as well. The explanation about why these vectors satisfy such properties can be found at the end of this subsection.

The Lorentz transformations connecting a given initial vector to a final one are referred to as \emph{Lorentz boost links}. Several transformations addressing this problem can be found in the literature (see, for example, \Mycitep{oziewicz2006lorentz}; \Mycitep{celakoska2015parameterization}). We have chosen the simplest option that fits our scheme, that is 
\[
\Lambda_{b}^{a}=\delta_{b}^{a}-\delta\Lambda_{b}%
^{a}~,%
\]
where
\[
\delta\Lambda_{b}^{a}=\frac{1}{-1+\frac{{\cal {\bar J}}_d {\cal {J}}^d}{{\cal {J}}%
		{\cal {\bar J}}}}
\left(  \frac{ {\cal {\bar J}}^a }{{\cal {\bar J}}}+ \frac{ {\cal {J}}^a }{{\cal {J}}} \right)
\left(  \frac{ {\cal {\bar J}}^b }{{\cal {\bar J}}}
+ \frac{ {\cal {J}}^b }{{\cal {J}}} \right)	 
+ \frac{ 2 {\cal {J}}^a {\cal {\bar J}}_b }{{\cal {J}} \, {\cal {\bar J}}}~,
\]
%
%$\norm{{\cal {J}}_d}$ 
being ${\cal {J}}=\sqrt{-{\cal {J}}_d {\cal {J}}^d}$ and ${\cal {\bar J}}=\sqrt{-{\cal {\bar J}}_d {\cal {\bar J}}^d}$ the norms of the respective moments.
%
%is a Lorentz boost transformation%
%(i.e., $\Lambda_{b}^{a} g^{bc} \Lambda_{c}^{d}%
%=g^{ad}$) 

We notice that this transformation is always well defined since, as we mentioned before, ${\cal{\bar J}}_d$ and ${\cal{J}}_d$ are timelike and future-directed. Therefore, $-1 \leq \frac{{\cal {\bar J}}_d {\cal {J}}^d}{{\cal {J}}{\cal {\bar J}}} < 0$, indicating that the denominators in the definition of $\delta\Lambda_{b}^{a}$ never go to zero.

This Lorentz boost link connects the two four-vectors as follows
\begin{equation}\label{eq_Lorertz_2}%
\frac{{\cal {\bar J}}^a}{{\cal {\bar J}}}
\Lambda_{a
}^{b}= \frac{{\cal {J}}^b}{{\cal {J}}} .
\end{equation}
We notice that, substituting Eq.~(\ref{eq_J_MC}) into this last expression, gives
\begin{equation}
{\cal {J}}^b=\sum_{k\in\Delta V} \frac{{N}_{k}}{\sqrt{\gamma}\Delta V} \\
\left(  \frac{{\cal {J}}}{{\cal {\bar J}}}
{p_k}^a \Lambda_{a
}^{b} \right), 
\end{equation}
%\end{widetext}
which leads to the desired result
\begin{equation}
{\cal {J}}^b = \sum_{k \in \Delta V} {N}_k \frac{\tilde{p}^k_b}{\sqrt{\gamma}\Delta V} ~~\textup{with}~~
\tilde{p}^k_b \equiv \frac{{\cal {J}}}{{\cal {\bar J}}}
{p_k}^a \Lambda_{a
}^{b} ~.
\end{equation}

The neutrino 4-momentum $\tilde{p}_{b}^{k}$ remains time-like (or null when $m_{\nu} = 0$) due to the fact that the Lorentz transformation $\Lambda_{b}^{a}$ preserves the norm of $p_{a}^{k}$. In simpler terms, we have applied a boost/rotation to the neutrino 4-momentum of all the packets within a given cell, ensuring that the \texttt{MC} moments ($\bar{E},\bar{F}_{i}$) precisely match the corresponding \texttt{GM} moments in that specific cell.

Returning to the discussion on the norms of ${\cal J}_a^{\texttt{MC}}$ and ${\cal J}_a^{\texttt{GM}}$. Let us assume a tiny non-zero neutrino mass  $m_{\nu}=\delta$, although negligible compared with the other scales in the problem. As indicated by Eq.~(\ref{eq_J_MC}), the direct consequence is that the 4-moment ${\cal J}_a^{\texttt{MC}}$ constitutes then a positive linear combination of future-directed timelike vectors . Consequently, ${\cal J}_a^{\texttt{MC}}$ must also be timelike and future-directed. To ensure the timelike and future-directed nature of ${\cal J}_a^{\texttt{M1}}$, we impose a consistency condition $F_i F^i \leq (1 - \delta) E^2$ (see \Mycitep{levermore84}). Finally, considering that ${\cal J}_a^{\texttt{GM}}$ (see Eq.~(\ref{eq_J_GM})) is a positive linear combination of ${\cal J}_a^{\texttt{M1}}$ and ${\cal J}_a^{\texttt{MC}}$, we can conclude that it is also timelike and future-directed.

%%%%%%%%%%%%%%%%%%%%%%%%%%%%%%%%%%%%%%%%%%%%%%%%%%%%%%%%%%%%%%%%%%%%%%%
\section{Numerical Tests}\label{sec4}

In this section we perform several stringent tests to validate our method and assess its accuracy. The numerical scheme employed for solving the \texttt{M1} formalism was presented in detail in \Mycite{paper_m1}. In summary, the evolution equations are evolved in time by using an Implicit-Explicit Runge-Kutta (IMEX) time integrator, fourth order accurate for the explicit part and second order for the stiff terms. The spatial discretization is based on a High-Resolution Shock Capturing (HRSC) scheme with Finite Volume reconstruction, designed to be asymptotically preserving in the diffusion equation limit (\Mycitep{rad2022}).
The numerical scheme employed to solve the \texttt{MC} formalism has been elaborated upon throughout the manuscript, and it is predominantly inspired by the ones presented in (\Mycitep{fou_moment}; \Mycitep{Foucart_2021}; \Mycitep{kawaguchi22}).

In the following we present the results of four distinct numerical solutions: \texttt{M1} represents the uncoupled (i.e., stand-alone) truncated moment approach, \texttt{MC} denotes the uncoupled Monte-Carlo method, \texttt{GM1} represents \texttt{M1} solutions, using the \texttt{GM} formalism, and \texttt{GMC} designates  \texttt{MC} solutions, using the \texttt{GM} formalism. These four solutions, each with its significant interest, will be subjected to thorough comparative analysis in the following numerical tests.

%%%%%%%%%%%%%%%%%%%%%%%%%%%%%%%%%%%%%%%%%%%%%%%%%%%%%%%%%%%%%%%%%%%%%%%		
\subsection{Diffusion in a Moving Medium Test}\label{T1}
\begin{figure}[t]%[h!]%[t!]%[htb!]
	\includegraphics[width=\columnwidth]{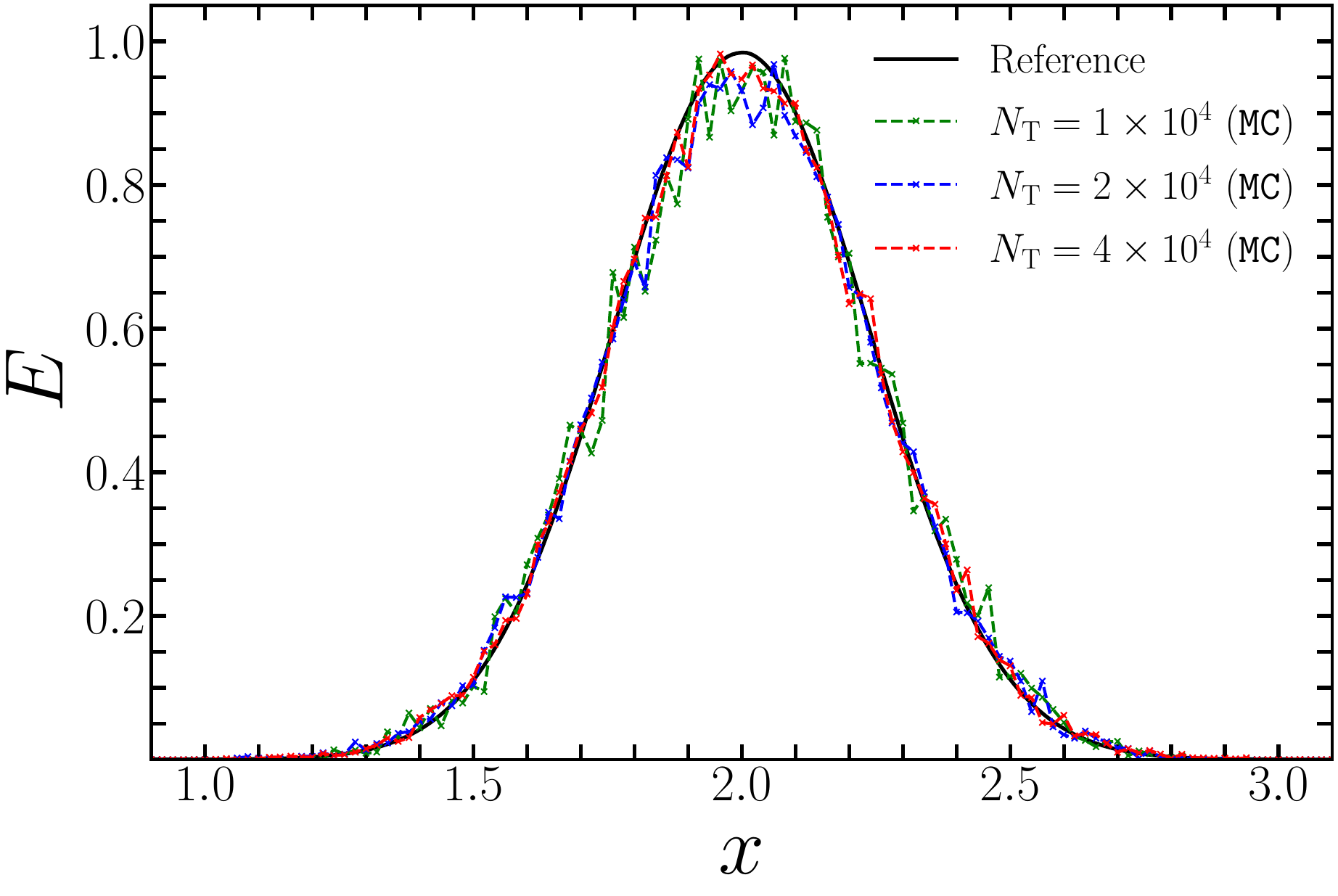}
	\caption{\emph{Diffusion in a Moving Medium Test}. 
		The profile of the energy density, reconstructed from the \texttt{MC} numerical solution, is displayed at time $t=4$ for different total number of packets $N_{\text{T}}$ by keeping the spatial resolution $\Delta x = 0.02$ fixed. As $N_{\text{T}}$ increases, the numerical solution tends to the semi-analytic (reference) one.}
	\label{T1_1}
\end{figure}
Diffusion in a moving medium is a challenging test, considered also in (\Mycitep{rad2022}; \Mycitep{paper_m1}; \Mycitep{musolino2023}; \Mycitep{cheong2023}),
which encapsulates most of the essential elements of the \texttt{M1} implementation. The test consists on a pulse of radiation energy density, characterized by a Gaussian profile, propagating within a moving medium which is dominated by scattering dynamics (i.e., a high value of the scattering opacity $\kappa_s$). The complete dynamics can be accurately captured only with the correct treatment of stiff source terms. 

The initial conditions at $t=0$ are specified as follows: $E( \mathbf{x}) = e^{-9x^2}$, $v^x(\mathbf{x}) = 0.5$, $\kappa_{s}(\mathbf{x}) =10^3$ and $\kappa_{a}(\mathbf{x}) = \eta(\mathbf{x})=0$.
The radiation fluxes, $F_{i}$, are initialized under the assumption that the radiation is fully trapped (i.e., $H^{a} = 0$). Using the relations between frames, this condition translates into
\begin{equation}
	J = \frac{3E}{4W^2 - 1} \,, \quad~~~~~ F_i = \frac{4}{3}JW^2v_i \,.
\end{equation}

In order to check the convergence of the numerical solutions we have considered three different spatial resolutions, corresponding to $\Delta x = (0.01,0.02,0.04$). For the evolution of the distribution function we have considered different fixed total number of packets $N_{\text{T}} = (0.5, 1 , 2 , 4) \times 10^4$.

\begin{figure}[t]%[h!]%[t!]%[htb!]
	\includegraphics[width=\columnwidth]{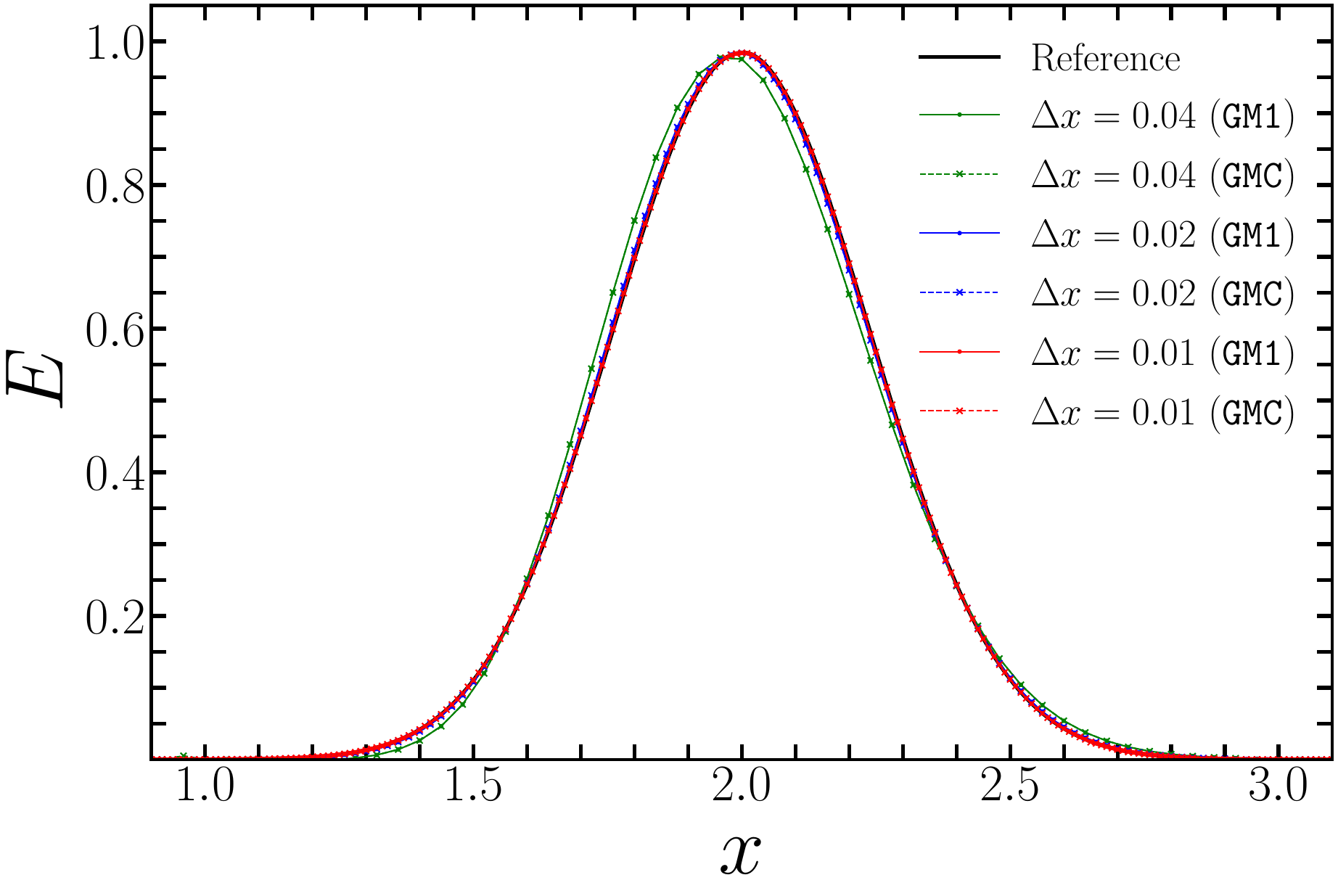}
	\caption{{\em Diffusion in a Moving Medium Test}.
	Comparison between the numerical solutions \texttt{GM1} and \texttt{GMC} and the semi-analytic solution for several grid resolutions $\Delta x$ with a number of packets $N_{\text{T}} = 2 \times 10^4$ fixed. As $\Delta x$ decreases, the numerical solution tends to the semi-analytic (reference) one.}
	\label{T1_2}
\end{figure}
Fig.~\ref{T1_1} illustrates the radiation energy density profile at time $t=4$, reconstructed from the \texttt{MC} solution, for various total numbers of packets $N_{\text{T}}=( 1, 2, 4 )\times 10^4$. The semi-analytic solution is also plotted for comparison purposes. A large number of packets has been employed to demonstrate that, as expected, with ``infinite" resolution (i.e., $N_{\text{T}} \rightarrow \infty$), the \texttt{MC} formalism converges to the true solution. However, the number of packets required for an accurate \texttt{MC} solution in high-scattering regions is unfeasible in realistic scenarios of neutron star mergers. 

This limitation is addressed by the \texttt{GM} scheme. In Fig.~\ref{T1_2}, the numerical solutions of \texttt{GM1} and \texttt{GMC} are plotted for different spatial resolutions $\Delta x = (0.01,0.02,0.04$), keeping $N_{\text{T}} = 2 \times 10^4$ fixed. As the entire domain of this test is optically thick, we expect an almost exact matching of the lowest moments from \texttt{M1} to \texttt{MC} with our choice of the function $h(\xi)$ (see Eq.~(\ref{h_xi}) and Fig.~\ref{closure_comparison}). Consequently, \texttt{GM1} and \texttt{GMC} are anticipated to precisely follow the \texttt{M1} behavior (for a comparison between the \texttt{M1} solution and the semi-analytic solution we refer to \Mycitep{paper_m1}). Indeed, as it is shown in Fig.~\ref{T1_2}, both \texttt{GM1} and \texttt{GMC} match perfectly the \texttt{M1} solution, approaching the semi-analytic one as the resolution increases. It is worth noting that, in such optically thick regimes, the \texttt{MC} method tends to be computationally slow and usually provides a poor solution. Therefore, by choosing $h(\xi)$ as specified, we can ensure that our \texttt{GM} method remains as accurate and efficient as the \texttt{M1} in this regime. %
\begin{figure}[t]%[t!]%[htb!]
	\includegraphics[width=\columnwidth]{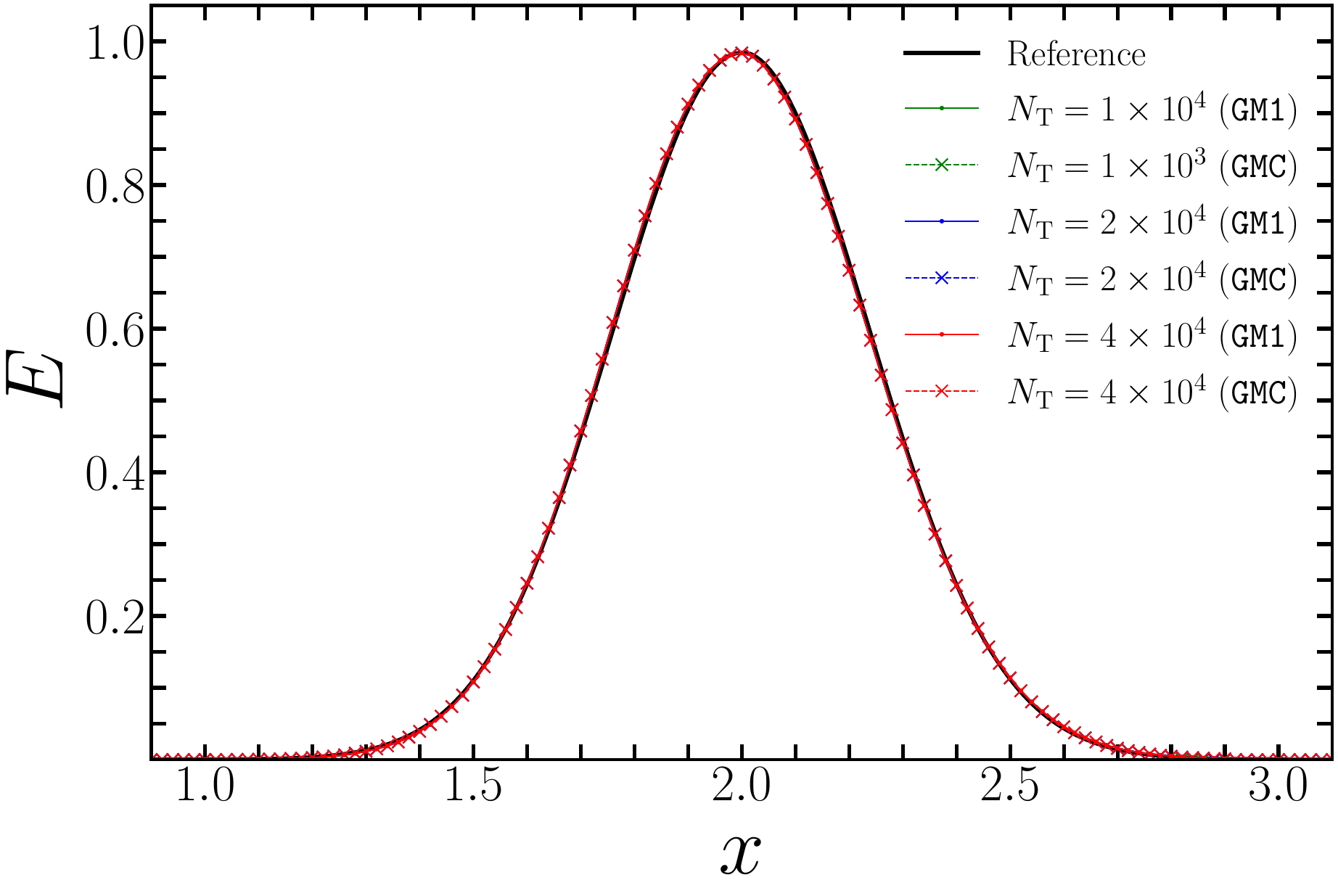}
	\caption{{\em Diffusion in a Moving Medium Test}.
		 Comparison between the numerical solutions \texttt{GM1} and \texttt{GMC} and the semi-analytic solution for various values of the total number of packets $N_{\text{T}}$ with a grid resolution $\Delta x = 0.02$ fixed. The observed perfect matching shows that the \texttt{GM} solutions do not depend on the total number of packets in optically thick regimes.}
	\label{T1_3}
\end{figure}%

These results are confirmed in Fig.~\ref{T1_3}, where the radiation energy density of the two \texttt{GM} numerical solutions is displayed by varying the total numbers of packets $N_{\text{T}} = (1, 2, 4) \times 10^4$ but keeping the spatial resolution $\Delta x = 0.02$ fixed. As anticipated, both \texttt{GM} solutions remains unchanged, overlapping almost exactly with the semi-analytical one. Consequently, within the \texttt{GM} scheme, increasing the number of packets is not necessary to enhance the numerical solution in optically thick regions.

\begin{figure}[t!]%[t!]%[htb!]
	\includegraphics[width=\columnwidth]{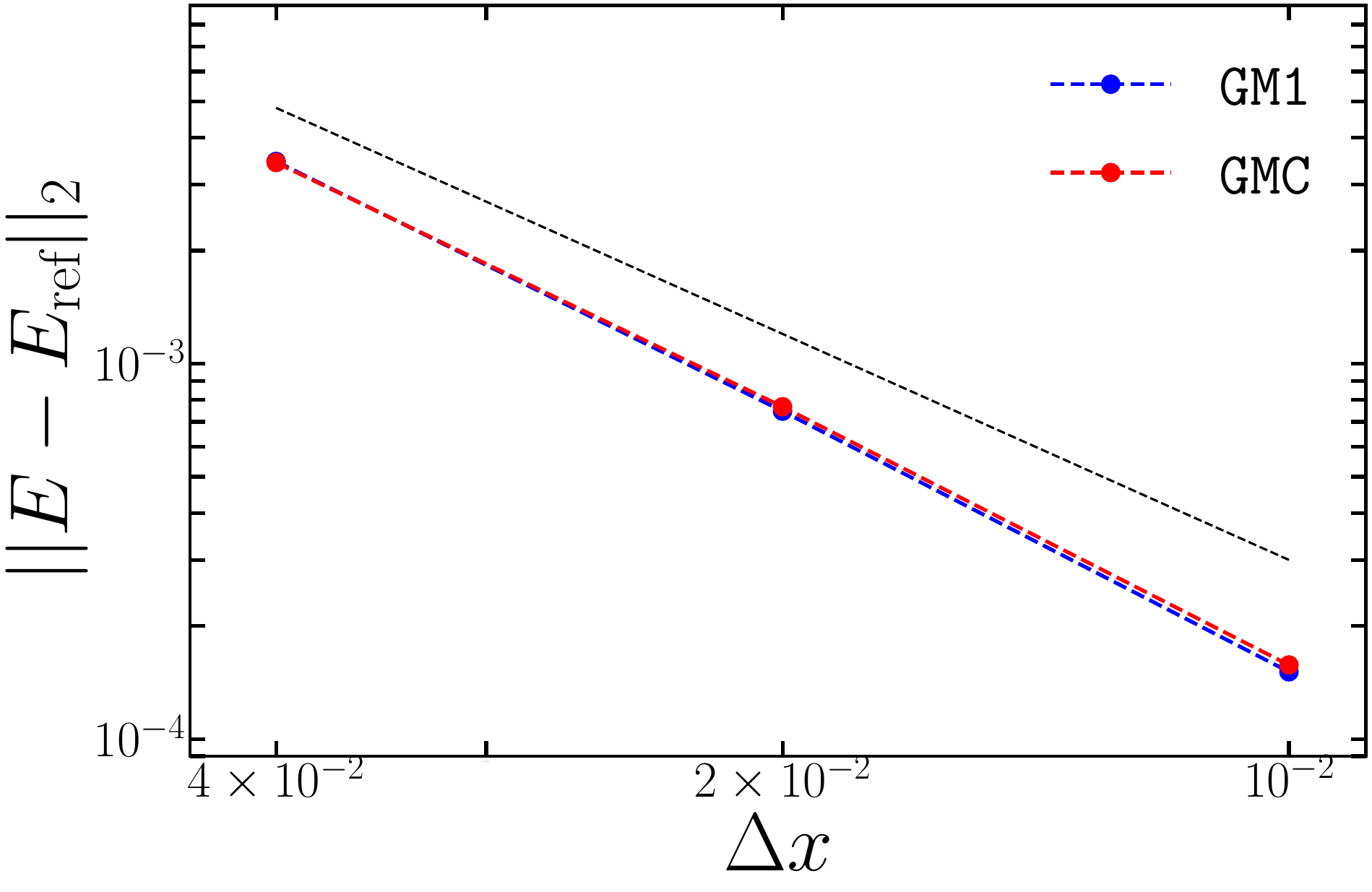}
	\caption{{\em Diffusion in a Moving Medium Test}. Convergence of the \texttt{GM1} and \texttt{GMC} solutions to the reference solution for a fixed number of total packets $N_{\text{T}}=2\times10^4$ by varying the grid resolution. We find an approximate second order convergence (dashed black line) for both \texttt{GM1} and \texttt{GMC} solutions.}
	\label{T1_4}
\end{figure}

A more quantitative analysis of the observed results can be obtained by performing suitable convergence tests of the error, which can be defined as the (norm of the) difference between the numerical and the semi-analytical solution. Two types of convergence can be examined: either varying the total numbers of packets while keeping the grid resolution fixed, or varying the grid resolution while keeping the total number of packets fixed. 
The latter one is represented in Fig.~\ref{T1_4} for a fixed number of total packets $N_{\text{T}}=2\times 10^4$. In this case, we find the expected second order of convergence of the \texttt{M1} scheme, as reported in \Mycite{paper_m1}. As mentioned earlier, both the \texttt{GM1} and \texttt{GMC} solutions closely follow the \texttt{M1} solution. 
There exists a very small difference in the error between the \texttt{GM1} and the \texttt{GMC}, originated by few cells which contain no packets and where the matching of the moments could not be performed.
\begin{figure}[t!]%[t!]%[htb!]
	\includegraphics[width=\columnwidth]{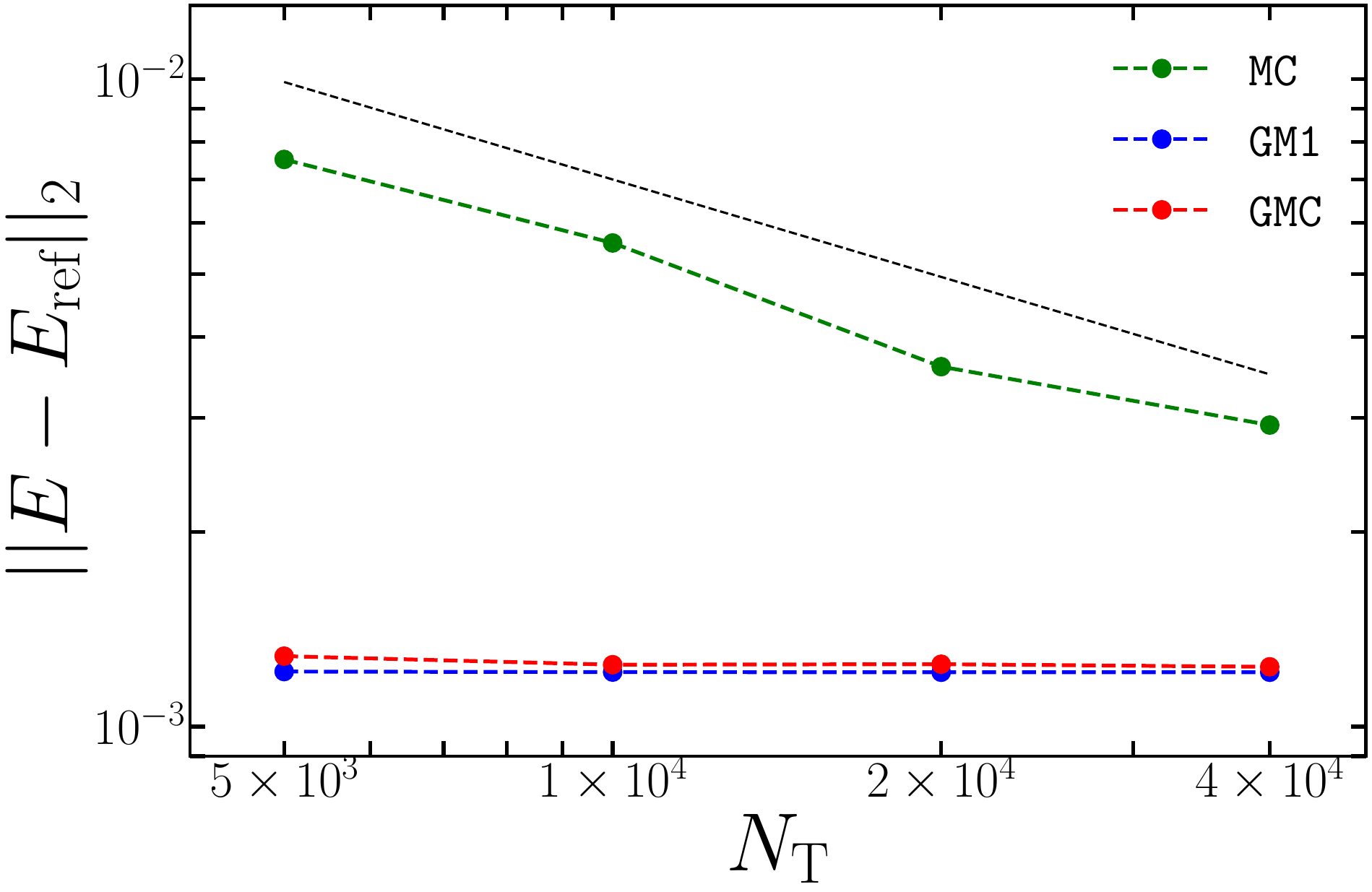}
	\caption{{\em Diffusion in a Moving Medium Test}. Convergence of the \texttt{GM1} and \texttt{GMC} solutions to the reference solution for a fixed grid resolution $\Delta x = 0.02$ by varying the number of total packets. We find the expected $N_{\text{T}}^{-1/2}$ convergence for the \texttt{MC} solution (dashed black line), but no dependence on $N_{\text{T}}$ for either the \texttt{GM1} or the \texttt{GMC} solution.}
	\label{T1_5}
\end{figure}%
The convergence keeping a fixed grid resolution $\Delta x = 0.02$ is  displayed in Fig.~\ref{T1_5}. Here, we find the anticipated \texttt{MC} convergence of $N_{\text{T}}^{-1/2}$. However, the error for both \texttt{GM1} and \texttt{GMC} is much smaller and exhibit no improvement when increasing the total number of packets. This lack of improvement is attributed to the accuracy dominance of \texttt{M1} over \texttt{MC} in the numerical solution for this test.
\begin{figure}[t] % Use [t] or [b] to control vertical alignment
	\centering
	\includegraphics[width=0.5\textwidth]{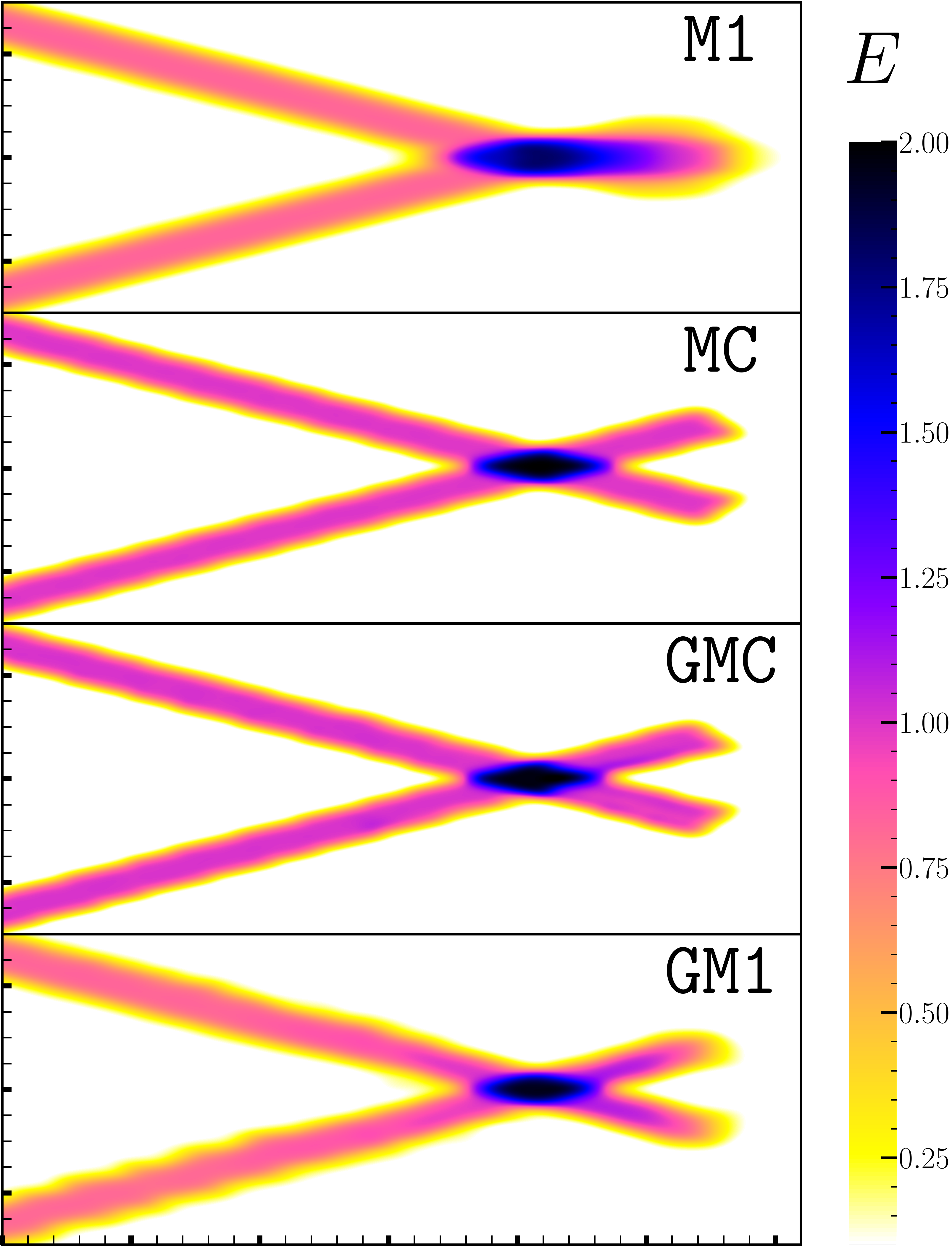}
	\caption{{\em Double Beam Test}. Numerical solutions for the (\texttt{M1}, \texttt{MC}, \texttt{GMC}, \texttt{GM1}) schemes are presented at the final time $t = 0.5$, using $\Delta x = 0.01$ and $N_{\text{T}} = 5 \times 10^6$ fixed. The behavior of \texttt{M1} reveals a breakdown in the optically thin limit when multiple  sources are present. In contrast, the (\texttt{MC}, \texttt{GMC}, \texttt{GM1}) solutions preserve the initial directions of the beams, enabling them to cross without interacting.}
	\label{T2_0}
\end{figure}
\begin{figure*}[t]%[t]%[h!]%[t!]%[htb!]
	\centering
	\includegraphics[width=2\columnwidth]{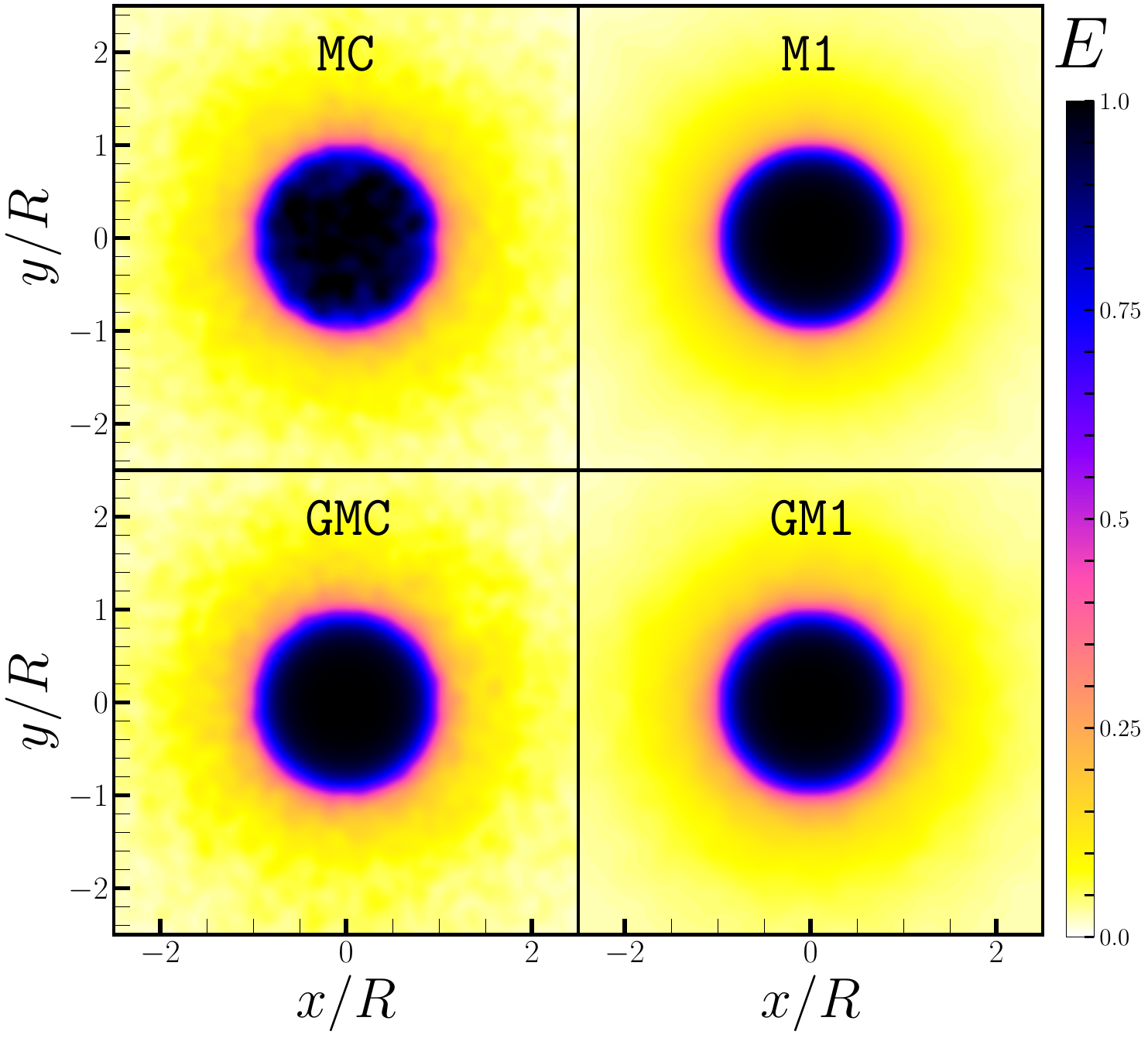}	\caption{{\em Radiating and Absorbing Sphere Test}. The energy density, computed with the (\texttt{MC}, \texttt{M1}, \texttt{GMC}, \texttt{GM1}) schemes using $\Delta x = 0.03$ and $N_{\text{tr}} = 80$, is displayed at the equatorial plane at time $t = 5$, when the solution has reached a steady state.}
	\label{T3_0}
\end{figure*}
%
%%%%%%%%%%%%%%%%%%%%%%%%%%%%%%%%%%%%%%%%%%%%%%%%%%%%%%%%%%%%%%%%%%%%%%%
\subsection{Double Beam Test}\label{T2}
This benchmark highlights the spectacular failure of the \texttt{M1} formalism in optically thin regions when multiple sources are present (see, e.g., \Mycitep{mck14}; \Mycitep{fou15}; \Mycitep{weih20}). In this two dimensional test, two radiation beams are injected into the domain.
If we consider the radiation to be neutrinos, one would expect the beams to follow straight paths, crossing without interaction. However, the \texttt{M1} scheme, treating radiation as a fluid, encounters challenges in this scenario since the closure is local and can only capture the physical behavior for a single source. On the other hand, the \texttt{MC} approach has no difficulties to find the correct physical solution.

In particular, we set the following initial conditions. We inject two beams of neutrinos from the left boundaries of the domain $(x, y) \in [-2, 2] \times [-4.5,4.5]$ with an angle of $60^{\circ}$ degrees between them. In this test we consider only a relatively high-resolution case, with a fixed spatial grid resolution $\Delta x = 0.01$ and total number of packets $N_{\text{T}} = 5 \times 10^6$. The simulation is performed with the four schemes (\texttt{M1}, \texttt{MC}, \texttt{GMC}, \texttt{GM1}) until the final time $t=0.5$, capturing the moment at which the beams have already intersected and continue along their trajectory.

The solutions of the energy density at this final time are presented in Fig.~\ref{T2_0}. As it was already mentioned, when two (or more) beams are present, the closure for the second moment in the \texttt{M1} formalism lacks sufficient information, causing the two beams to merge into a single one propagating along the average of the original directions.
Notice that, although the distribution function contains all the information about possible directions of propagation, the \texttt{M1} formalism retains only a single averaged direction: momenta in opposite directions cancel each other out, resulting in a loss of information regarding the original momentum distribution (\Mycitep{weih20}). In principle, the lost information could potentially be recovered by employing higher moments.
In clear contrast, in the \texttt{MC} scheme the two beams intersect seamlessly without losing energy density or modifying their trajectories. A similar outcome, though not as pristine, is observed for the \texttt{GMC} and \texttt{GM1} numerical solutions. This outcome highlights a significant advantage of our \texttt{GM} formalism, which  demonstrates an accurate handling of the solution in optically thin regions.

%
%%%%%%%%%%%%%%%%%%%%%%%%%%%%%%%%%%%%%%%%%%%%%%%%%%%%%%%%%%%%%%%%%%%%%%%	
\subsection{Radiating and Absorbing Sphere Test}\label{T3}
A more challenging problem, which includes both the optically thick and thin regimes, is given by an a homogeneous radiating and absorbing sphere. This test, whose analytical solution is known, has been widely discussed in astrophysical literature (see, e.g., \Mycitep{smit97}; \Mycitep{pons00}; \Mycitep{rampp02}; \Mycitep{rad13}; \Mycitep{anninos20}, \Mycitep{weih20}, \Mycitep{chan20}; \Mycitep{rad2022}; \Mycitep{paper_m1}; \Mycitep{musolino2023}; \Mycitep{cheong2023}), since it can be interpreted as a highly simplified model for an isolated, radiating neutron star.

The main configuration involves a static (i.e., $v^i(\mathbf{x})=0$), spherically symmetric, homogeneous sphere of radius $R$ with a constant energy density. In this idealized case, the only neutrino-matter interaction process allowed is the isotropic thermal absorption and emission.
\begin{equation}
	\begin{aligned}
		\kappa_{a}(\mathbf{x}, \mathbf{y}, \mathbf{z}) &= \eta(\mathbf{x},\mathbf{y}, \mathbf{z}) =			
		\begin{cases}
			10 & \text{for } r \leq R~, \\
			0 & \text{for } r > R~,
		\end{cases} \\
		\kappa_{s}(\mathbf{x},\mathbf{y}, \mathbf{z}) &= \mathbf{0}~.
	\end{aligned}
\end{equation}
where $r= \sqrt{x^2 + y^2 + z^2}$ and we have chosen $R=0.5$. 
Our 3D computational Cartesian domain is a cube of dimensions $[-2,2]^3$, discretized with different spatial grid resolution resolutions $\Delta x = (0.03, 0.06, 0.12)$. In this specific scenario, characterized by a non-zero emissivity, the simulation cost of the Monte-Carlo scheme is governed by the free parameter $N_{\text{tr}}$, related to the neutrino packets generated in a grid cell (see Eq.~(\ref{Ntr})). In this test, we have explored various values of $N_{\text{tr}} = (20, 40, 80)$.
\begin{figure}[t!]%[t!]%[htb!]
\includegraphics[width=\columnwidth]{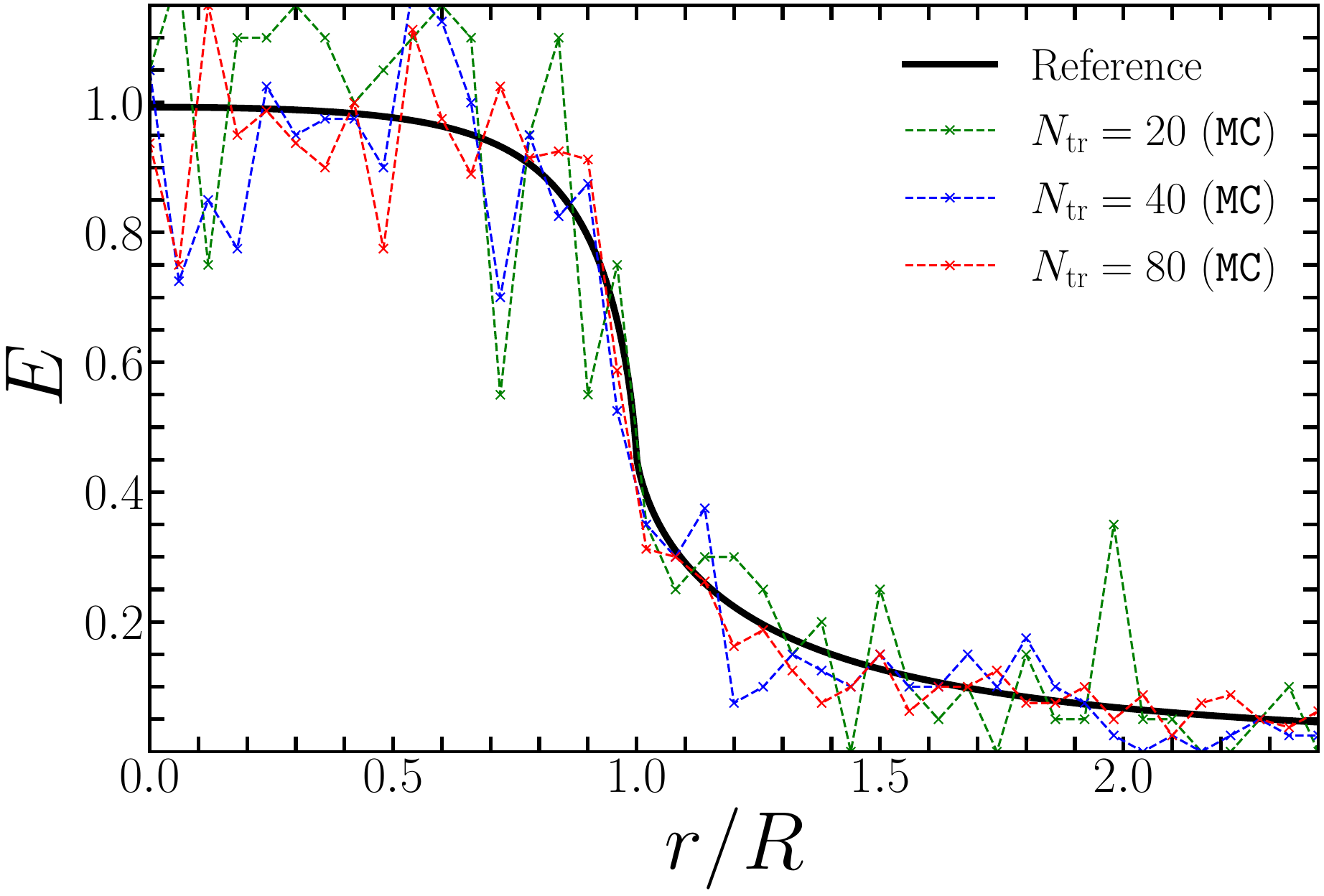}
\caption{{\em Radiating and Absorbing Sphere Test}. Comparison between the \texttt{MC} numerical and the exact solutions for various values of the number of packets $N_{\text{tr}}$ with a fixed spatial resolution $\Delta x = 0.03$. The statistical noise of the numerical solution decreases as the number of packets increases. }\label{T3_1}
\end{figure}
\begin{figure}[t!]%[t!]%[htb!]
	\includegraphics[width=\columnwidth]{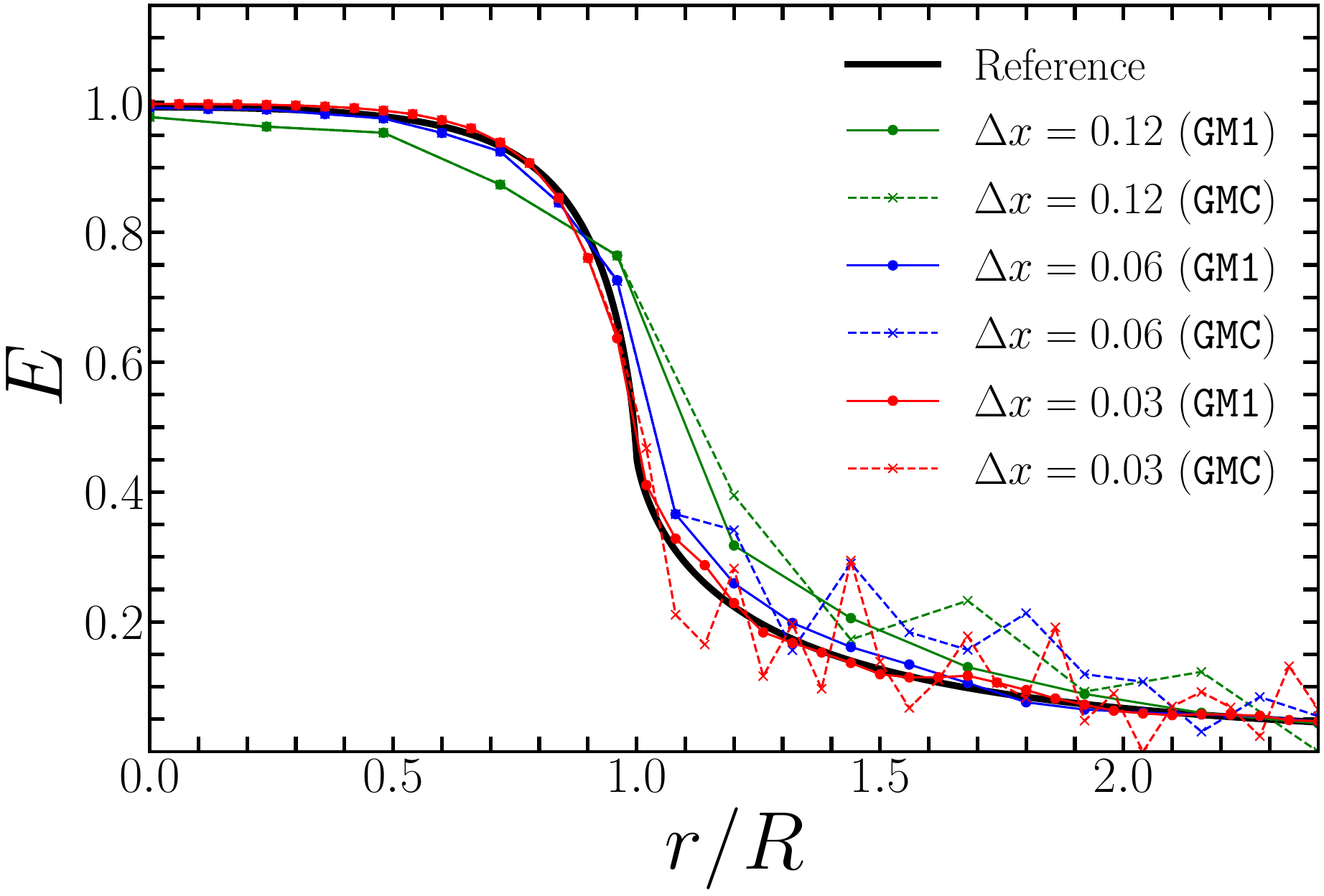}
	\caption{{\em Radiating and Absorbing Sphere Test}. Comparison between the numerical solutions \texttt{GM1} and \texttt{GMC} and the analytic solution for various grid resolutions with a fixed number of packets $N_{\text{tr}} = 40$. The \texttt{GM1} clearly converges to the exact solution. However, in the case of \texttt{GMC}, while there is clearly convergence in the interior of the star, the presence of statistical noise remains fairly consistent over all grid resolutions.}\label{T3_2}
\end{figure}
\begin{figure}[h!]%[t!]%[htb!]
	\includegraphics[width=\columnwidth]{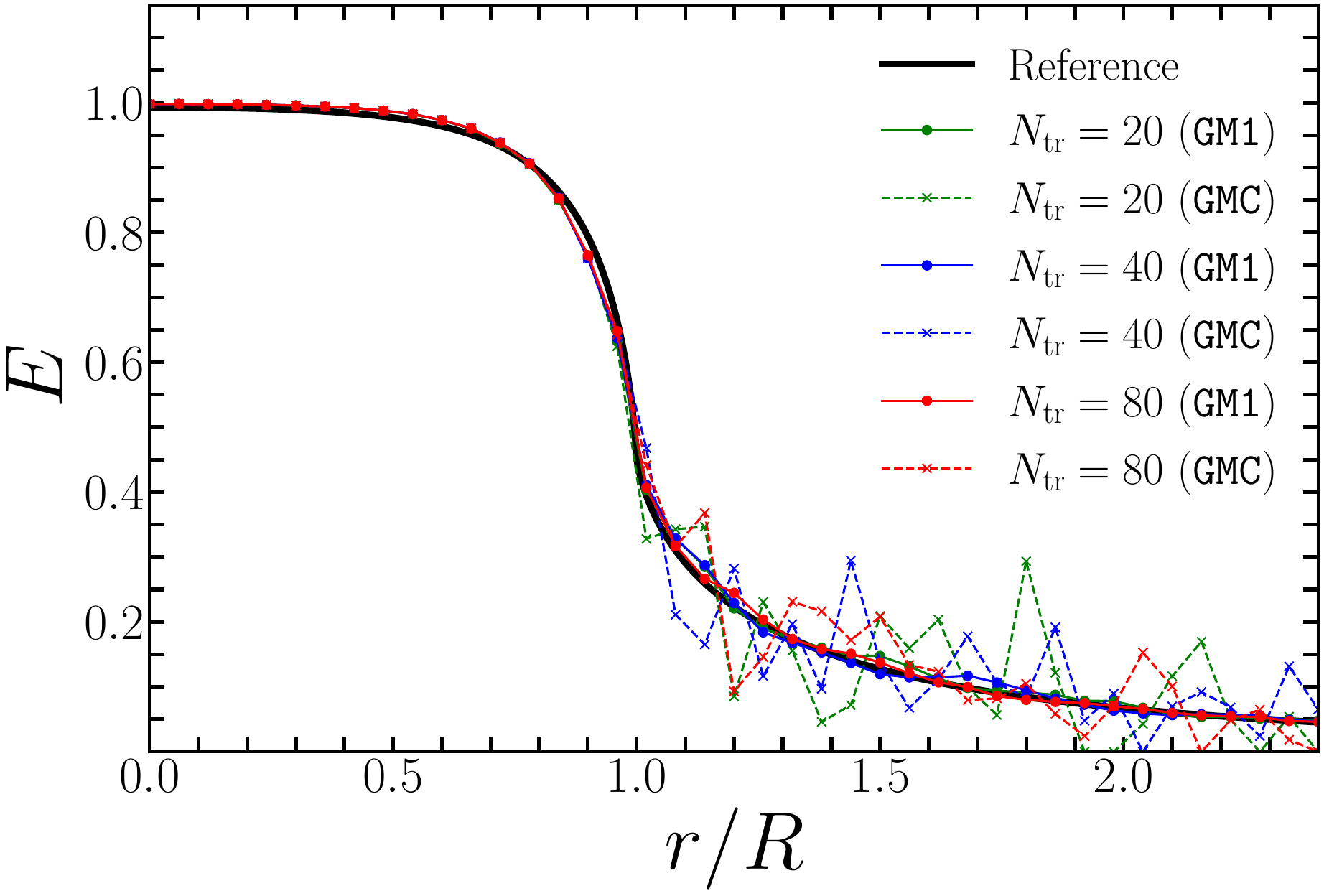}
	\caption{{\em Radiating and Absorbing Sphere Test}. Comparison between the numerical solutions for \texttt{GM1} and \texttt{GMC} and the analytic one for various values of the number of packets while keeping a fixed grid resolution $\Delta x = 0.03$. As in Fig. \ref{T3_1}, the statistical noise decreases for both numerical solutions as the number of packets increases.}\label{T3_3}
\end{figure}
\begin{figure}[t!]%[t!]%[htb!]
	\includegraphics[width=\columnwidth]{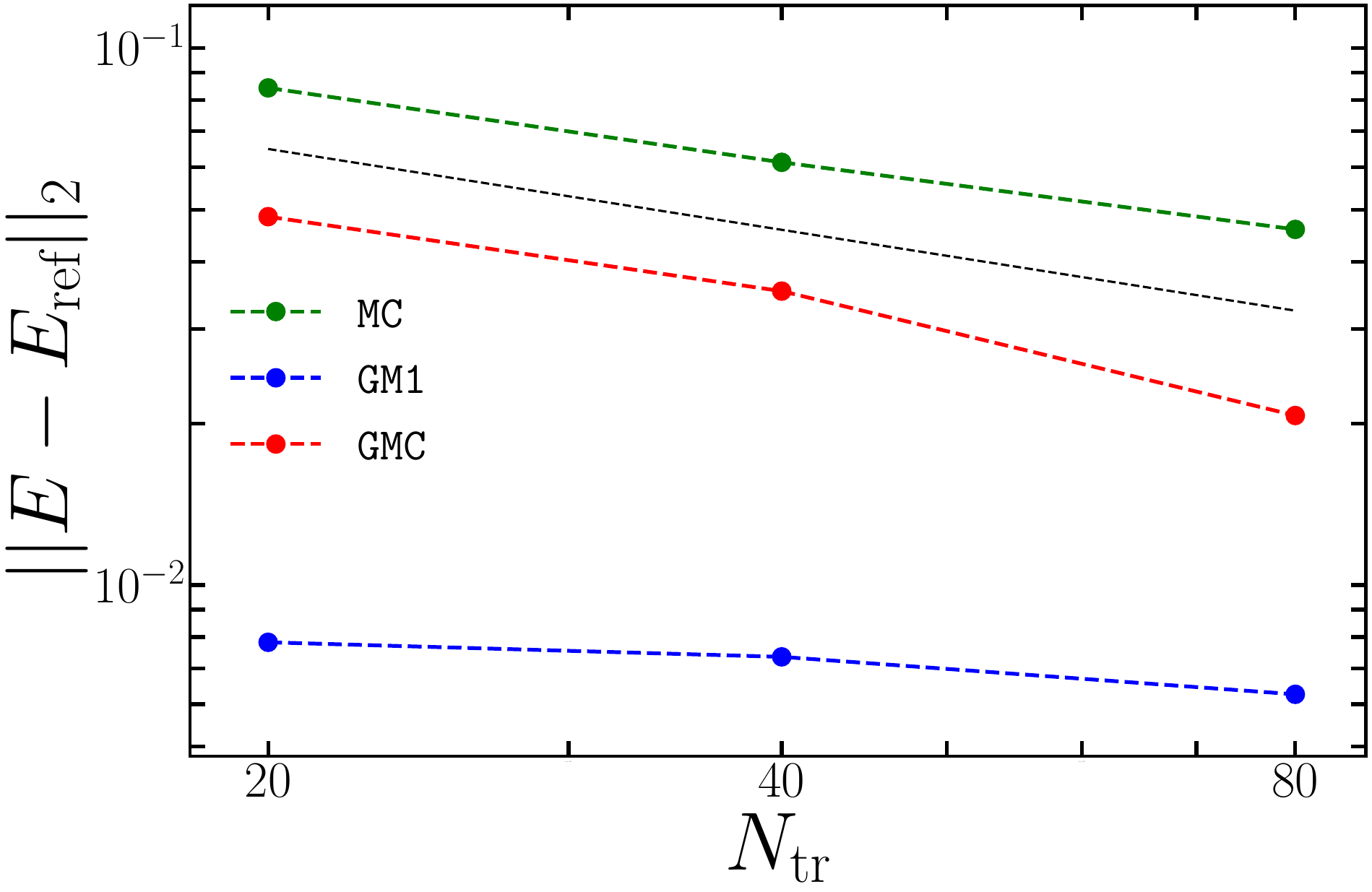}
	\caption{{\em Radiating and Absorbing Sphere Test}. Convergence of the (\texttt{MC}, \texttt{GM1},  \texttt{GMC}) solutions for a fixed grid resolution $\Delta x = 0.03$ by varying the number of packets. We find the expected $N_{\text{T}}^{-1/2}$ convergence (dashed black line) of the \texttt{MC} method except for the \texttt{GM1} solution, which presents already a very small error.}\label{T3_4}
\end{figure}
\begin{figure}[t!]%[t!]%[htb!]
	\includegraphics[width=\columnwidth]{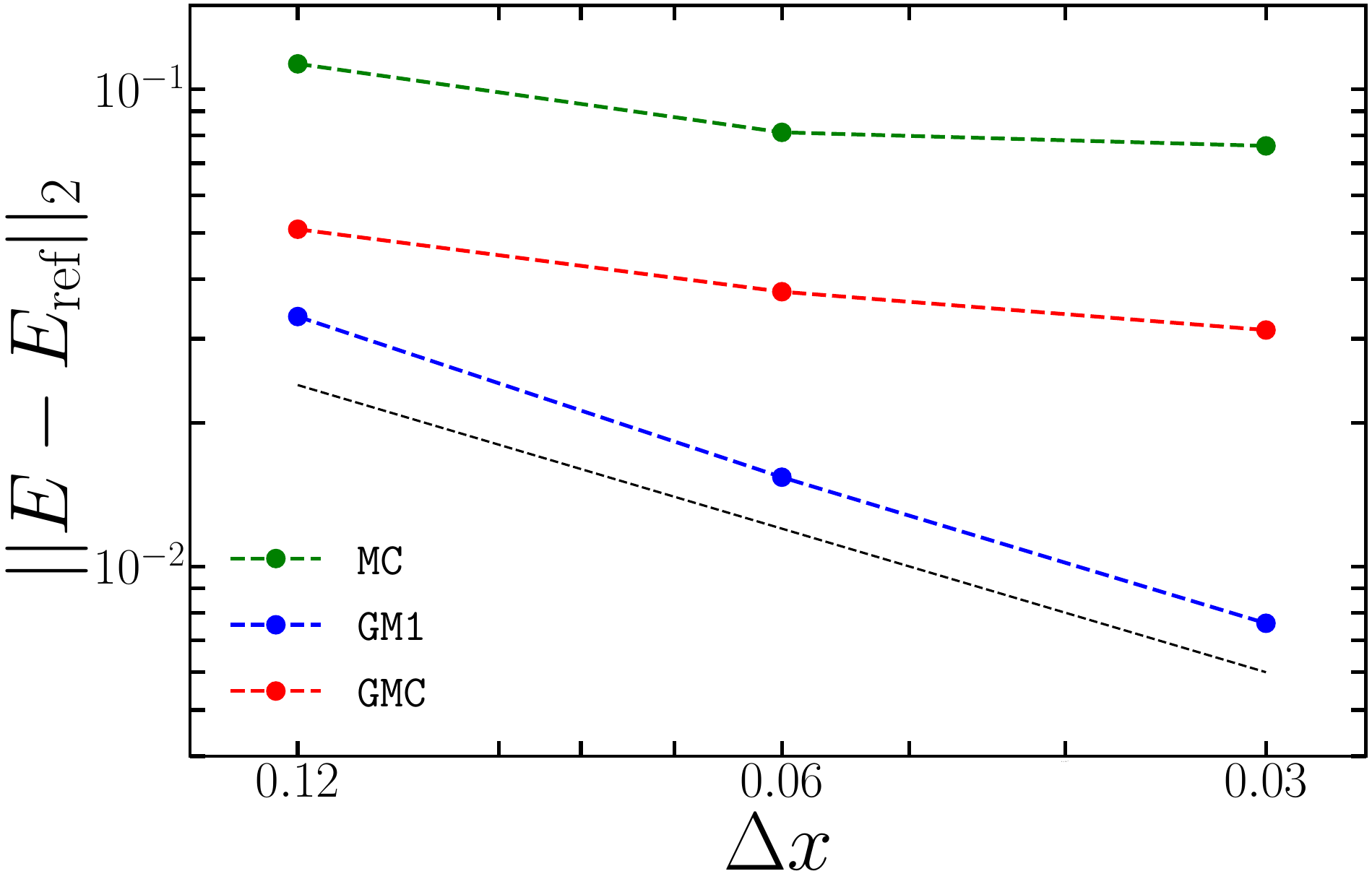}
	\caption{{\em Radiating and Absorbing Sphere Test}. Convergence of the (\texttt{MC}, \texttt{GM1}, \texttt{GMC}) solutions for a fixed number of packets $N_{\text{tr}}=40$ while varying the  grid resolutions. We find an approximate first order convergence (dashed black line) for the \texttt{GM1} but only a moderate improvement for the \texttt{MC} or \texttt{GMC} solutions, which only converge as the number of packets increases.}\label{T3_5}
\end{figure}

Fig.~\ref{T3_0} displays the radiation energy density in the equatorial plane of the four considered schemes (\texttt{MC}, \texttt{M1}, \texttt{GMC}, \texttt{GM1}) at time $t = 5$, when the solution has reached a steady state. These results, obtained using $\Delta x = 0.03$ and $N_{\text{tr}} = 80$, are consistent with our theoretical expectations. All methods produce a qualitative similar solution, with \texttt{MC} and \texttt{GMC} showing oscillations throughout the domain due to statistical noise (i.e., which would diminish with a greater number of packets),  while the \texttt{M1} and \texttt{GM1} exhibit much smoother profiles. In particular, the \texttt{GMC} effectively resolves accurately the interior of the sphere, where we observe a full matching from the \texttt{M1} solution onto the \texttt{MC} one, as expected in optically thick regions. In the exterior of the sphere, some oscillations appear due to the stochastic nature of the \texttt{MC} method, which is dominant in the optically thin region. The \texttt{GM1} solution matches the \texttt{M1} in the interior of the sphere, and it smoother than the \texttt{GMC} in the exterior. This positive outcome confirms that the \texttt{GM} approach retains the advantages of evolving a truncated moment scheme while incorporating new features that allow to handle accurately also the optically thin limit.

In order to assess quantitatively the accuracy of our methods, the radial profile of the energy density is compared against the analytical (reference) solution. The \texttt{MC} solution is presented in Fig.~\ref{T3_1} for various numbers of packets corresponding to $N_{\text{tr}} = (20, 40, 80)$, while keeping the spatial resolution $\Delta x = 0.03$ fixed. The inherent statistical noise of the \texttt{MC} is evident for all points in the domain, yet the solution progressively converges to the analytic one as the number of packets is increased. This test highlights one of the limitations of the \texttt{MC} scheme, namely its lower convergence rate. On the other hand, the solution of the \texttt{M1} formalism displays a second order convergence when appropriate numerical schemes are employed (we refer to \Mycitep{paper_m1} for details).

We now turn on the results of our \texttt{GM} formalism. In Fig.~\ref{T3_2}, we compare the numerical solutions of \texttt{GM1} and \texttt{GMC} with the analytic solution for different spatial resolutions $\Delta x = (0.03,0.06,0.12)$, while keeping the number of total packets approximately constant setting $N_{\text{tr}}=40$. Given the optically thick regime found in the star's interior, the \texttt{GMC} solution exhibits no statistical noise in this region, unlike the outer region where oscillations are present. Interestingly, these oscillations are mostly suppresed in the \texttt{GM1} solution. The average nature of the truncated moments formalism and the HRSC schemes employed for the evolution of the moments are probably the cause of this suppresion. Finally, notice that the \texttt{GM1} solution converges to the analytical one as the grid spacing decreases. However, the \texttt{GMC} solution only converges clearly in the interior of the sphere. Outside there remains a statistical noise that does not depend on the grid resolution. In Fig.~\ref{T3_3}, we compare the numerical solutions of \texttt{GM1} and \texttt{GMC} versus the analytic solution for different choices of the number of packets  $N_{\text{tr}}=(20,40,80)$ keeping the spatial resolution $\Delta x = 0.03$ fixed. As expected, the statistical noise outside the sphere is reduced as the number of packets increases.

Convergence tests are performed by varying the number of total packets while keeping the spatial resolution fixed and vice versa. The results of the former analysis are presented in Fig.~\ref{T3_4}, where the error in the numerical solution is displayed as a function of the total number of packets for a fixed spatial resolution $\Delta x = 0.03$. Notably, the \texttt{GM1} solution exhibits minimal improvement with increasing number of packets, given the dominance of the \texttt{M1} solution in the interior, which already has negligible error, and the smoothness of the solution in the exterior. On the other hand, both \texttt{GMC} and \texttt{MC} demonstrates an approximate $1/2$ convergence, attributed to the suppression of statistical noise in the exterior of the star as the total number of packets increases.
The convergence of the solutions as a function of the spatial grid resolution while keeping a fixed number of packets $N_{\text{tr}} = 40$ is displayed in Fig.~\ref{T3_5}. Here, the \texttt{GM1} solution shows a first-order convergence, while the \texttt{MC} and \texttt{GMC} solutions displays only a marginal improvement. This moderate improvement is  mainly due to the projection of the \texttt{M1} solution inside the star, while the solution in the exterior is still dominated by statistical noise which only decreases by increasing the number of packets. This result confirms that the \texttt{GM} formalism only converges to the exact solution for infinite grid resolution and infinite number of packets.
\begin{figure*}[t!]%[t]%[h!]%[t!]%[htb!]
	\centering
	\includegraphics[width=2\columnwidth]{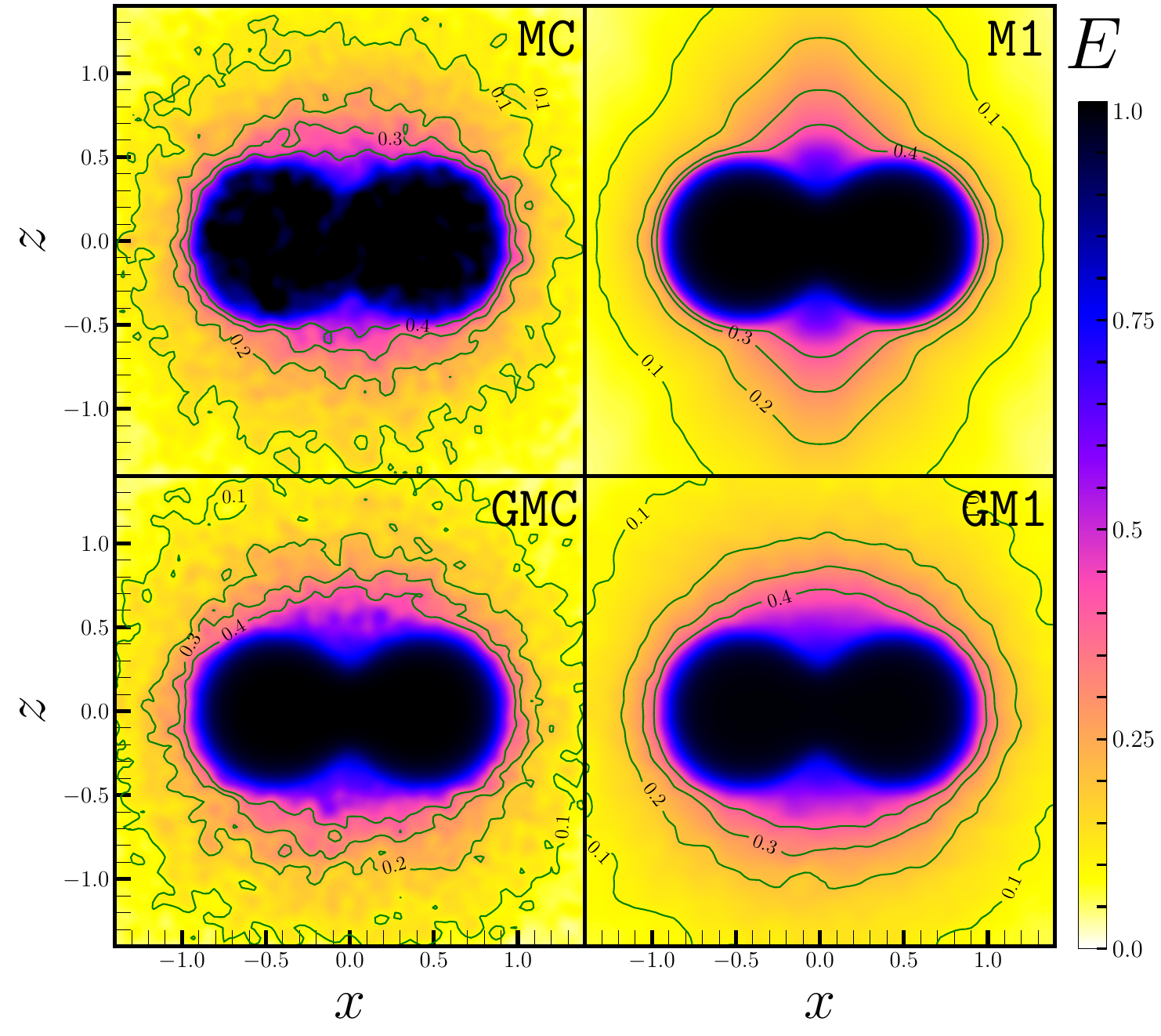}	\caption{{\em Radiating and Absorbing Torus Test}. The energy density, computed with the (\texttt{MC}, \texttt{M1}, \texttt{GMC}, \texttt{GM1}) schemes using $\Delta x = 0.03$ and $N_{\text{tr}} = 40$, is displayed at the meridional plane at time $t = 5$, when the solution has reached a steady state.	
}
	\label{T4_0}
\end{figure*}
%
%%%%%%%%%%%%%%%%%%%%%%%%%%%%%%%%%%%%%%%%%%%%%%%%%%%%%%%%%%%%%%%%%%%%%%%
\subsection{Radiating and Absorbing Torus Test}\label{T4}
In our last 3D problem we simulate a scenario reminiscent of the one discussed in Sec.~\ref{T2}, but without so many symmetries. The astrophysical motivation behind this configuration is to reproduce a simple model of the remnant resulting from a neutron star merger. Instead of an absorbing and radiating sphere, the geometry of the system is better approximated by a torus, which can be described by the following parametric equations in Cartesian coordinates:
\begin{align*}
	x(u, v) &= (R + r \cos v) \cos u~, \\
	y(u, v) &= (R + r \cos v) \sin u~, \\
	z(u, v) &= r \sin v~.
\end{align*}
The coordinate position on the torus is determined by the parameters $(u,v) \in [0, 2\pi)$, while that the major and minor radius $(R,r)$ define the type of torus. For our specific configuration, we have chosen a self-intersecting spindle torus (i.e., $r>R$) with $r = 0.5$ and $R = 0.45$~. 
Inside the torus, we set again high emissivity and absorption opacity $\eta = \kappa_a = 10$, such that neutrinos in this region are in equilibrium with the fluid, representing the optically thick regime. Outside the torus, both the emissivity and absorption opacity are set to zero $\eta = \kappa_a = 0$, modeling an optically thin medium. As the simulation begins, neutrinos emitted from the interior propagate to the exterior . Due to the geometry of the self-intersecting spindle torus, beams will collide along the symmetry axis (i.e., along the $z=0$ axis),  resembling the collision of multiple beams of radiation. The simulation is performed in a 3D Cartesian cubic domain with dimensions $[-3, 3]^3$, employing a fixed spatial resolution $\Delta x = 0.03$ and number of packets $N_{\text{tr}} = 40$.

In Fig.~\ref{T4_0} we present the final time $t=5$, when the system has reached a steady state, for the four schemes considered (\texttt{MC}, \texttt{M1}, \texttt{GMC}, \texttt{GM1}). The plot displays a 2D slice along the meridional direction (i.e., $y=0$ plane). Similar to the previous test (Sec.~\ref{T3}), the \texttt{MC} solution appears less smooth due to its intrinsic statistical noise. The crucial distinction between the exact methods and the \texttt{M1} becomes evident when looking along the $z$-axis.  In the \texttt{M1}, a shock forms where beams collide, representing the previously discussed failure of this approach in the optically thin limit. On the other hand, in the cases of (\texttt{MC}, \texttt{GM1}, \texttt{GMC}), the solutions do not have a large-energy-density region along this axis caused by the artificial collision of radiation. Crucially, our \texttt{GM} formalism, particularly the \texttt{GM1} solution, exhibits a solution with smoothness comparable to that of the \texttt{M1}, but without the presence of shocks. This result confirms and validates the accuracy of the \texttt{GM} method both in the optically thick and thin regimes.
%
%%%%%%%%%%%%%%%%%%%%%%%%%%%%%%%%%%%%%%%%%%%%%%%%%%%%%%%%%%%%%%%%%%%%%%%
\section{\textbf{Conclusions}}\label{S_conclusions}
Modeling accurately the intricate physical scenarios within neutron star mergers, especially during the post-merger phase, demands sophisticated numerical simulations. Neutrinos, being crucial contributors to the dynamics and thermodynamics of these events, require a specialized treatment. Here we have introduced a novel approach, the \textbf{\emph{Guided Moments}} (\texttt{GM}) formalism, to achieve an accurate and efficient full-neutrino transport treatment in complex environments like  neutron star mergers and core-collapse supernovae. This formalism efficiently combines the advantages of the truncated moments scheme (\texttt{M1}) and Monte-Carlo (\texttt{MC}) based methods, providing a robust solution that addresses the strengths and weaknesses of each method. 

%To understand the intrinsics of our method, we first discuss how the \texttt{M1} and \texttt{MC} schemes work in the optically thick and thin regimes. 
One of the key concepts in the \texttt{GM} formalism is to compute the closure of the \texttt{M1} evolution equations (i.e., the second moment) by using information from the \texttt{MC} solution. 
In the optically thick limit, this closure is analytical and provides already a very accurate and efficient solution. In the same regime, however, the \texttt{MC} scheme has an opposite behavior and faces with two challenges: the continuous emission and absorption of neutrino packets and the always present statistical noise associated to stochastic processes. To mitigate these issues, in our \texttt{GM} scheme, we use information from the \texttt{M1} solution to modify the neutrino distribution function such that the \texttt{MC} lowest moments matches the ones evolved by the \texttt{M1} formalism.
On the other hand, the \texttt{M1} closure in the optically thin limit is not known for multiple sources, so here our \texttt{GM} formalism takes  advantage of the cost-effectiveness of the \texttt{MC} scheme to compute the exact closure in this regime. 

These previous points and the deep discussion included in this paper, allow us to say that the \texttt{GM} formalism not only accurately captures the optically thick limit through the exact \texttt{M1} closure, but also effectively resolves the optically thin limit, a known challenge for the \texttt{M1} approach but accurately handled by \texttt{MC} methods. The resulting scheme outperforms both the \texttt{M1} and \texttt{MC} approaches, providing a comprehensive and accurate solution in both regimes. Although we have focused on the pressure tensor closure, our method likely also improves the energy closure. Studying this issue is beyond the scope of the paper, but it will be thoroughly discussed in future work involving astrophysical scenarios.

The detailed exposition of the \texttt{GM} formalism, its formulation, and implementation, along with a thorough comparison against  \texttt{M1} and \texttt{MC} methods across various test problems, demonstrates the efficacy of the proposed approach. The computational cost of evolving the \texttt{MC} solution in optically thick regions can be substantial in real simulations. This issue can be mitigated within the \texttt{GM} scheme by effectively limiting the emissivities and opacities only in \texttt{MC} scheme for these regions, thereby reducing the overall computational cost. As there is a complete matching of the lowest moments of \texttt{M1} and those of \texttt{MC}, there should not be any degradation of the accuracy in the \texttt{GM} solution.

Another potential improvement is related to modifying the matching function $h(\xi)$ (see Eq.~(\ref{h_xi}) and Fig.~\ref{closure_comparison}), which determines the regime in which we will match the lowest moments calculated with the \texttt{M1} solution with those obtained from the \texttt{MC} scheme. While the presented test problems show promising behavior, fine-tuning might be necessary in more realistic simulations and probably will require a more careful exploration. 
%
%%%%%%%%%%%%%%%%%%%%%%%%%%%%%APENDICES%%%%%%%%%%%%%%%%
\appendix	
%%%%%%%%%%%%%%%%%%%%%%%%%%%%%%%%%%%%%%%%%%
\section{Matching the neutrino number density}\label{N_app}
%%%%%%%%%%%%%%%%%%%%%%%%%%%%%%%%%%%%%%%%%%%

Recent extensions of the \texttt{M1} formalism (see, e.g., \Mycitep{Foucart_2016}; \Mycitep{rad2022}; \Mycitep{paper_m1}) also evolve the number density of neutrinos. Notice that, without evolving the number density, the \texttt{M1} formalism with the grey approximation does not conserve accurately the lepton number (\Mycitep{Foucart_2016}). Although this issue might be not so dramatic in the \texttt{GM} approach, where there is a better estimate of the neutrino energy spectrum, it might  still be problematic. To this aim, for each neutrino species one can introduce a neutrino number current $N^a$ following a conservation equation
\begin{equation}
	\nabla_a N^a = \sqrt{-g} {\cal C} =  \sqrt{-g} (\eta^0 - \kappa_a^0 n)~,
\end{equation}
where $n = -N^a u_a$ is the neutrino density in the fluid frame and $(\kappa_a^0, \eta^0)$ are the neutrino number absorption and emission coefficients, also to be computed from the fluid state and the information in the EoS tables.

Assuming that the neutrino number density and the radiation flux are aligned, this equation can be written in the $3+1$ decomposition as
\begin{widetext}
	\begin{eqnarray}
		\partial_t (\sqrt{\gamma} {N}) 
		&+& \partial_k \left[\sqrt{\gamma} \left(- W \beta^k + \alpha W v^k + \alpha \frac{H^k}{J}\right) \frac{{N}}{\Gamma} \right] = \alpha \sqrt{\gamma} \left( \eta^0 - \kappa_a^0 \frac{N}{\Gamma} \right)~,
	\end{eqnarray}
\end{widetext}
where $N= -n_a N^a = n \Gamma$ is the neutrino density in the inertial frame and
\begin{eqnarray}
	\Gamma = W - \frac{1}{J} H^a n_a = 
	W \left( \frac{E - F_a v^a}{J} \right)~.
\end{eqnarray}
Within this new equation, one can estimate dynamically the average energy of the neutrinos $\epsilon_{\nu}$, since in the fluid frame the relation $J \approx n<\epsilon_{\nu}>$ is approximately satistied.

In the guided moments formalism we are matching the lowest moments, but it might be desirable also to match the neutrino number density if it is being evolved also in the \texttt{M1}.

It is straightforward to show that, in the discrete packet distribution, the neutrino number density and the lowest moments, measured in the grid frame, can be computed just as
\begin{equation}
	{\bar N} = \sum_{k \in \Delta V}  \frac{N_k}{\sqrt{\gamma}\Delta V}
	~,~~~
	{\cal {\bar J}}_a = \sum_{k \in \Delta V} N_k \frac{p^k_a}{\sqrt{\gamma}\Delta V}~.
\end{equation}

The first step is to match the density of neutrino number ${\bar N}$ to $N$, a task that requires a modification of the weights $N_k$. This can be achieved by performing the following simple renormalization:
\begin{equation}
	\tilde{N}_{k} \equiv \frac{N}{{\bar N}}N_{k} \rightarrow
	N = \sum_{k \in \Delta V}  \frac{\tilde{N}_k}{\sqrt{\gamma}\Delta V}~,
\end{equation}
which imply that the projection of the stress-energy tensor needs to be computed now as
\begin{equation}
	{\cal {\bar J}}_a = \sum_{k \in \Delta V} {\tilde N}_k \frac{p^k_a}{\sqrt{\gamma}\Delta V}~.
\end{equation}
Basically, the interpretation of this new relation for the moments is that by changing the number of neutrinos in a cell, automatically the energy and flux densities change in that cell.

The rest of the procedure for matching moments remains the same, such that the final result
\begin{equation}
	{\cal {J}}^b = \sum_{k \in \Delta V} {\tilde N}_k \frac{\tilde{p}^k_b}{\sqrt{\gamma}\Delta V} ~,
\end{equation}
involves now two transformations: changing $N_k \rightarrow \tilde{N}_{k}$ (i.e., modifying the weights such that the density of neutrinos are equal) and $p^k_b \rightarrow \tilde{p}^k_b$ (i.e., modifying the neutrino 4-momentum such that the lowest moments match).

%%%%%%%%%%%%%%%%%%%%%%%%%%%%%%%%%
\section{Tetrad}\label{tetradMC}
%%%%%%%%%%%%%%%%%%%%%%%%%%%%%%%%%

Two special observers will play an important role in the description of our neutrino transport algorithm: inertial observers, whose timeline is tangent to $n^a$, and co-moving observers, whose
timeline is tangent to $u^a$. 

Our numerical grid is discretized in the spatial coordinates $x^i$. We will refer
to the coordinates $(t, x^i )$ as the inertial or grid frame, where the line element is
\begin{equation}
	ds^2 = g_{ab} dx^{a} dx^{b} ~.
\end{equation}
We also define the coordinates of the fluid rest frame $(t',x^{i'})$, which are defined at a point such that 
\begin{equation}
	ds^2 = \eta_{a'b'} dx^{a'} dx^{b'}~,
\end{equation}
with $\eta_{a'b'}$ the Minkowski metric, and $(t')^a = u^a$. We construct these local coordinates from an orthonormal tetrad $e^{(c')}_a$, with 
\begin{eqnarray}
	\hat{e}^a_{(t')} &=& u^a ~,\\
	g^{ab} \hat{e}_a^{(c')} \hat{e}_b^{(d')}&=& \eta^{c'd'}~.
\end{eqnarray}
The three other components of the tetrad are obtained by applying Gramm-Schmidt’s algorithm to the three vectors $V^a_{(i)} = \delta^a_i$ (i = 1, 2, 3). The orthonormal tetrad $\hat{e}_{(c')}^a$, and the corresponding one-forms $\hat{e}^{(c')}_a$, are precomputed and stored for each grid cell at each timestep, and can be used to easily perform transformations from the fluid rest frame coordinates to grid coordinates (and vice-versa) by simple matrix-vector multiplication.

In order to convert the 4-momentum from the lab frame ($\mathcal{S}$) to the fluid-rest-frame or comoving frame ($\mathcal{S'}$) we need to find a basis that relates $p^{c}$ and $p^{c'}$. This will be necessary when initializing the packets (set random moments), after scattering (redraw momentum) and when packets are emitted (new initialization). In these simulations, the momentum is draw in an spherical uniform distribution, i.e.,
\begin{equation}
	p^{c'} = \nu (1, \sin\theta \cos \phi, \sin\theta \sin \phi, \cos \theta)~.
\end{equation}
We draw $\cos(\theta)$ from a uniform distribution in $[-1, 1]$ and $\phi$ from a uniform distribution in $[0, 2 \pi]$. The 4-momentum of neutrinos in grid coordinates can then be computed using the transformation
\begin{eqnarray}
	p^t &=& \hat{e}^t_{(c')}p^{c'}~, \\
	p_i &=& g_{ia}\hat{e}^{a}_{(c')}p^{c'} = \delta_{c' d'}\hat{e}^{(c')}_{i}p^{d'}~.
\end{eqnarray}
%
%%%%%%%%%%%%%%%%%%%%%%%%%%%%%%%%%%%%%%%%%%%%%%%%%%%%%%%%%%%%%%%%%%%%%%%
\section*{\textbf{Acknowledgments}}
We thank Federico Carrasco, Lorenzo Pareschi and David Radice for suggestions and clarifications on the subjects of this work. 
MRI is grateful for the hospitality and stimulating discussions during his visit at  Stony Brook's Physics and Astronomy department and at the Institute for Gravitation and the Cosmos (Penn State University).
CP acknowledges the hospitality at the Institute for Pure \& Applied Mathematics (IPAM) through the Long Program ``Mathematical and Computational Challenges in the Era of Gravitational Wave Astronomy", where this project was initiated.
MRI thanks financial support PRE2020-094166 by MCIN/AEI/PID2019-110301GB-I00 and by ``FSE invierte en tu futuro''.
This work was supported by the Grants PID2022-138963NB-I00 and PID2019-110301GB-I00 funded by MCIN/AEI/10.13039/501100011033 and by ``ERDF A way of making Europe".%
%%%%%%%%%%%%%%%%%%%%%%%%%%%%%REFERENCES%%%%%%%%%%%%%%%%
%\bibliographystyle{unsrtnat}
\bibliography{refs}
%%%%%%%%%%%%%%%%%%%%%%%%%%%%%%%%%%%%%%%%%%%%%%%%%%%%%%%
\end{document}